\documentclass[12pt]{article} 

\usepackage{color}
\definecolor{darkblue}{rgb}{0.1,0.1,.7}
\usepackage[colorlinks, linkcolor=darkblue, citecolor=darkblue, urlcolor=darkblue, linktocpage]{hyperref} 
\usepackage[]{amsmath}
\usepackage[]{graphicx}
\usepackage[]{latexsym}
\usepackage{geometry}
\usepackage[margin=10pt,font=small,labelfont=bf]{caption}
\usepackage{amscd}
\usepackage{bm}
\usepackage{upgreek}
\usepackage[square, comma, sort&compress,numbers]{natbib}
\usepackage[all,cmtip]{xy}
\numberwithin{equation}{section}
\newcommand{\reef}[1]{(\ref{#1})}
\def\beq{\begin{equation}} 
\def\eeq{\end{equation}} 
\def\beqrr{\begin{array}} 
\def\eeqrr{\end{array}} 
\def\beqa{\begin{eqnarray*}} 
\def\eeqa{\end{eqnarray*}} 
 
\font\mybb=msbm10 at 12pt 
\def\bb#1{\hbox{\mybb#1}} 
 
\def\bR {\bb{R}}

\def\bM {\bb{M}}

\def\b1 {\bb{1}} 

\def\del {\partial}

\def\lesssim{\mathrel{\hbox{\rlap{\hbox{\lower4pt\hbox{$\sim$}}}\hbox{$<$}}}} 
\def\gtrsim{\mathrel{\hbox{\rlap{\hbox{\lower4pt\hbox{$\sim$}}}\hbox{$>$}}}}

\newcommand{\hhref}[1]{\href{http://arxiv.org/abs/#1}{arXiv:#1}}
\newcommand{\CG}{{$SO(d+1,1)$}}
\newcommand{\ES}{{$\bM^{d+2}$}}
\newcommand{\PS}{{$\bR^{d}$}}
\numberwithin{equation}{section}
\newcommand{\proof}{\vspace{0.2em}\footnotesize}
\def\eps{\varepsilon}

\begin{document} 

\vspace*{-.6in} \thispagestyle{empty}
\begin{flushright}
LPTENS--11/22\\
NSF-KITP-11-128\\
\end{flushright}
\vspace{.2in} {\Large
\begin{center}
{\bf Spinning Conformal Correlators}
\end{center}}
\vspace{.2in}
\begin{center}
Miguel S. Costa$^{a}$, Jo\~ao Penedones$^{b}$,  David Poland$^{c}$, and Slava Rychkov$^{d,e}$
\\
\vspace{.3in} 
\emph{${}^a\,$
Centro de F\'\i sica do Porto and 
Departamento de F\'\i sica e Astronomia\\
Faculdade de Ci\^encias da Universidade do Porto\\
Rua do Campo Alegre 687,
4169--007 Porto, Portugal}
\\
\vspace{.2in} 
\emph{${}^b\,$Perimeter Institute for Theoretical Physics\\ Waterloo, Ontario N2L 2Y5, Canada}
\\
\vspace{.2in} 
\emph{${}^c\,$Jefferson Physical Laboratory, Harvard University\\Cambridge, Massachusetts 02138, USA}
\\
\vspace{.2in} 
\emph{${}^d\,$Laboratoire de Physique Th\'{e}orique, \'{E}cole Normale Sup\'{e}rieure,\\
and Facult\'{e} de Physique, Universit\'{e} Pierre et Marie Curie (Paris VI)}
\\
\vspace{.2in} 
\emph{${}^e\,$KITP, University of California, Santa Barbara, USA}

\end{center}

\vspace{.3in}

\begin{abstract}
We develop the embedding formalism for conformal field theories, aimed 
at doing computations with symmetric traceless operators of arbitrary spin.
We use an index-free notation where tensors are encoded by polynomials
in auxiliary polarization vectors.  The efficiency of the formalism is demonstrated 
by computing the tensor structures allowed in $n$-point conformal correlation functions
of tensors operators.  Constraints due to tensor conservation also take a simple form in this
formalism.  Finally, we obtain a perfect match between the number of independent tensor 
structures of conformal correlators in $d$ dimensions and the number of independent 
structures in scattering amplitudes of spinning particles in $(d+1)$-dimensional Minkowski space.
\end{abstract}
\vskip 1cm \hspace{0.7cm} July 2011

\newpage

\setcounter{page}{1}

\tableofcontents

\break

\section{Introduction}

One hardly needs to stress the importance of Conformal Field Theories (CFT) in theoretical physics.
In two dimensions, many exactly solvable models exist, thanks to the infinite dimensional extension of the global conformal group,
the Virasoro algebra. Unfortunately, in three dimensions or higher, no equally efficient general approaches are known at present.

One approach which holds some promise is the `conformal bootstrap' \cite{FGG-paper,pol}, which tries to solve or constrain a higher-dimensional CFT
by imposing the Operator Product Expansion (OPE) associativity. The efficiency of this method has been demonstrated in several recent applications \cite{us,joao,david}.
However, so far this approach has been limited to the study of four-point functions of scalar operators.
It is of great interest to extend this technique to other operators like the stress-energy tensor or global symmetry currents.
This could provide very general constraints for any CFT or for CFTs with a given global symmetry. 
In this paper, we give the first step towards this goal by developing an efficient language to deal with primary tensor operators in CFT.
Basically, our formalism makes CFT computations with tensor fields as easy as computations with scalars.
In an upcoming paper~\cite{Costa:2011dw}, we shall use this formalism to obtain conformal blocks for four-point functions of tensor operators.

Another motivation for this work is the recently found analogy between CFT correlation functions written in the Mellin representation and scattering amplitudes~\cite{MackMellin,joaoMellin}.\footnote{See also~\cite{Fitzpatrick:2010zm} for a connection between CFT anomalous dimensions and scattering amplitudes.}
This analogy has been explored in detail in the case of CFT correlators defined holographically by Witten diagrams of scalar field theories in AdS 
\cite{joaoMellin,Fitzpatrick:2011ia,Paulos:2011ie}.  It would be very interesting to find a generalization to correlators of tensor operators.
The first steps towards this goal were given in \cite{Fitzpatrick:2011ia,Paulos:2011ie}.
It is natural to expect that such a generalization could lead to recursion relations for the computation of stress-energy tensor correlators in CFTs with AdS gravity duals,\footnote{See \cite{Suvrat} for a proposal based on momentum space correlators.}
 similar to the BCFW recursion relations for scattering amplitudes \cite{BCFW}.
More generally, one might hope to use this analogy to translate all the powerful methods for the computation of scattering amplitudes to CFT correlation functions (at least, for CFTs with a weakly coupled AdS dual).
We believe the formalism described in this paper to deal with tensor operators will also be useful in this context.  

In this paper, we test the analogy between $d$-dimensional conformal correlators and $(d+1)$-dimensional scattering amplitudes at the level of counting independent coupling constants. More precisely, we show that the number of tensor structures for three point correlators of tensor operators is equal to the number of tensor structures for three particle S-matrix elements in one higher dimension.
AdS/CFT provides a natural map from S-matrix elements of the bulk theory to correlators of the boundary CFT.
The idea is to define the correlator by the contact Witten diagram with local interaction vertex associated with the scattering amplitude.
This map can be used to obtain CFT $n$-point correlators from analytical $n$-particle S-matrix elements (contact interactions). However, for $n>3$, the scattering amplitudes can have poles associated with particle exchange diagrams. In this case, some similarity seems to persist but it is not obvious how to define an explicit map.

\textbf{Structure of the paper:} The paper is built upon the embedding space formalism, which we review in section \ref{sec:emb}. In this formalism \cite{Dirac,Mack:1969rr,Boulware,FGG-paper,FGG-book,CCP,Weinberg}, correlators in Euclidean $d$-dimensional  space are uplifted to homogeneous functions on the lightcone of $(d+2)$-dimensional Minkowski spacetime, where the conformal group acts as the Lorentz group. This goes a long way towards simplifying CFT computations, but in the case of tensor fields it still falls short of our needs. In section \ref{sec:tens}, we develop a version of the embedding formalism which encodes the index structure of the tensor operators in polynomials of a  `polarization vector' in $(d+2)$-dimensions. In section \ref{sec:corr}, we use the new \emph{index-free} formalism to compute constraints from conformal symmetry on correlators (3-, 4- and $n$-point functions) of tensor operators of arbitrary spin. We are able to rederive in a simplified and explicit way a number of known results, and to get some new ones. In section \ref{sec:cons} we show how to implement constraints on correlation functions of conserved tensors in our language.  In section \ref{sec:S} we discuss a rule which allows to count conformal $n$-point functions in terms of on-shell scattering amplitudes of higher spin massive fields  in $(d+1)$-dimensions and, in case of conserved tensors, massless fields. For the case of three-point functions of conserved operators with spin $l_i$ in dimension $d\ge 4$, this gives the number of allowed tensor structures to be $1+{\rm min} (l_1,l_2,l_3)$. Section \ref{sec:concl} gives a summary of the new algorithm for dealing with CFT correlation functions  and concludes.

\section{Embedding Formalism}
\label{sec:emb}

In this paper we consider CFT in $d\ge 3$ Euclidean dimensions, so that the conformal group is \CG.
All of our equations can be Wick-rotated to the Minkowski signature, paying attention to the $i\epsilon$ prescription. We assume that the reader is familiar with the basics of the theory, see e.g.~\cite{DF}, chapter 4. 
As is well known, conformal symmetry imposes strong constraints on the correlation functions of primary operators in the theory.  These constraints are relatively easy to work out for primary scalars, but they become less transparent for primary fields of nonzero spin. In this section we will develop the `embedding formalism' which makes the nonzero spin case easier.  The formalism has been applied on and off since the early CFT days \cite{Boulware,FGG-book}. We will take as a starting point a version used recently in \cite{CCP} (see also \cite{Weinberg} for a recent discussion).\footnote{Additional work using six-dimensional field equations to describe four-dimensional theories has been done, e.g., in~\cite{Bars}.}

The basic idea, due to Dirac \cite{Dirac}, is that the natural habitat for the conformal group {\CG} is the \emph{embedding space} \ES, where it can be realized as the group of linear isometries. Thus, conformal symmetry constraints should become as trivial as Lorentz symmetry constraints,  provided 
all CFT fields can somehow be lifted to \ES. 
The lift is accomplished via a sort of stereographic projection. First, a point  $x\in\text{\PS}$ is put in correspondence with a null ray in \ES\ consisting of the vectors
\beq
P^A  = \lambda \left(1,x^2 ,x^a\right)\,,\qquad \lambda\in \bR\,,
\label{lambda}
\eeq
where we use light cone coordinates
 \begin{equation}
P^A = \left(P^+,P^-,P^a\right)\,,
\end{equation}
with metric given by\footnote{Here $\delta_{ab}\to\eta_{ab}$ when Wick-rotating to the Minkowski spacetime signature.}
\begin{equation}
P\cdot P  \equiv \eta_{AB} \,P^A P^B = - P^+P^- + \delta_{ab }\,P^a P^b\,.
\end{equation}
Here and below, we use capital letters to denote embedding space (\ES) quantities and lower case letters to denote physical space (\PS) quantities.

\begin{figure}[htbp]
\begin{center}
\includegraphics[scale=0.4]{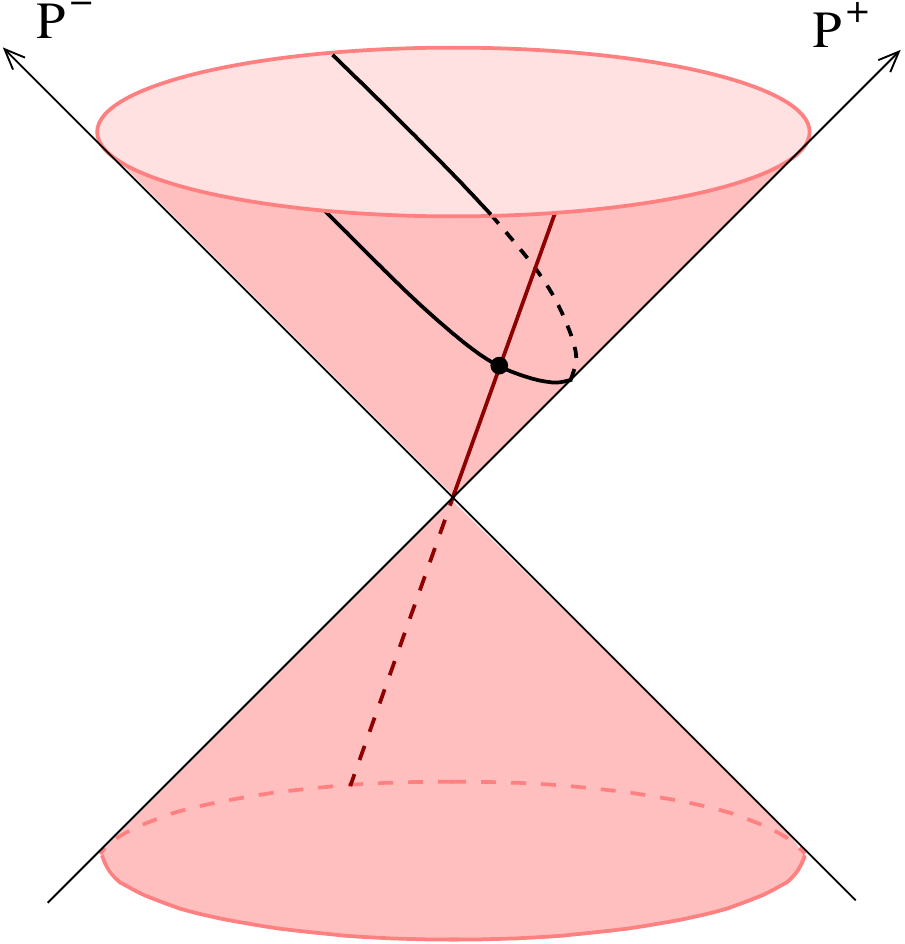}
\caption{Light cone in the embedding space; light rays are in one-to-one correspondence with physical space points. The Poincar\'e section of the cone 
is also shown.}
\label{Cone}
\end{center}
\end{figure}

Now, a linear \CG\ transformation of \ES\ will map null rays into null rays, and via Eq.~(\ref{lambda}) this defines a map of the physical space \PS\ into itself, which turns out to be a conformal transformation in the usual sense. Moreover, every conformal transformation can be realized this way \cite{Dirac}. 

Next we should establish the correspondence between fields on \PS\ and \ES, which is done as follows. Consider a field $F_{A_1\ldots A_l}(P)$, a tensor of \CG, with the following properties:
\begin{enumerate}
\item Defined on the cone $P^2=0$.
\item Homogeneous of degree $-\Delta$:  $F_{A_1\ldots A_l}(\lambda P)=\lambda^{-\Delta}F_{A_1\ldots A_l}(P)$, $\lambda>0$.
\item Symmetric and traceless.
\item Transverse: $(P\cdot F)_{A_2\ldots A_l}\equiv P^A F_{AA_2\ldots A_l}=0$. 
\end{enumerate} 
Notice that all these conditions are manifestly \CG-invariant. 
Because of homogeneity, $F$ is known everywhere on the cone once it is known on the \emph{Poincar\'e section},\footnote{Other sections of the cone could be useful to study CFT on curved, conformally flat, backgrounds.}
\beq
P^A_x=(1,x^2,x^a)\,,\quad\quad x\in \text{\PS}\,,
 \label{section}
 \eeq
whose vectors are in one-to-one correspondence with the points of \PS. Projecting $F$ to the Poincar\'e section defines a symmetric tensor field on 
\PS:\footnote{Here and below, we omit the dependence of $P_x$ on $x$ in ${\partial P}/{\partial x}$, to avoid cluttering.}
\beq
f_{a_1\ldots a_l}(x)= \frac{\partial P^{A_1}}{\partial x^{a_1}}\dots  \frac{\partial P^{A_l}}{\partial x^{a_l}} F_{A_1\dots A_l}(P_x)\,.  
\label{proj}
\eeq
This operation has two important properties. First, any tensor proportional to $P_A$ projects to zero. We will call such \CG\ tensors \emph{pure gauge} \cite{Boulware}. It is not difficult to show that if two symmetric transverse tensors $F$ and $F'$ project to the same $f$, then they differ by pure gauge (this is valid point by point on the Poincar\'e section). 

Second, the projected tensor is traceless, as long as $F$ is traceless and transverse. This follows from the identity
\beq
K^{AB}\equiv \delta^{ab} \frac{\partial P^{A}}{\partial x^{a}}  
  \frac{\partial P^{B}}{\partial x^{b}} =\eta^{AB}+P_x^A\bar{P}^{B}+P_x^B\bar{P}^{A}\,, \qquad \bar{P}=(0,2,0)\,,
  \label{id1}
\eeq
which is easily verified by using the explicit form of the projection matrices:
\beq
 \frac{\partial P^{A}}{\partial x^{c}}=(0,2x_c,\delta^a_c)\,.
 \label{expl}
 \eeq

Given that any conformal transformation can be realized as an \CG\ rotation, and that $F$ transforms as a tensor of \CG, it makes sense to ask how $f$ defined by (\ref{proj}) transforms under the conformal group. It can be shown \cite{FGG-book, Weinberg}\footnote{Ref.\ \cite{FGG-book} imposes a divergence-free condition to fix the pure gauge terms in $F$, which leads to unnecessary complications.} that this transformation is exactly that of a spin $l$ symmetric traceless primary field of dimension $\Delta$.  This is actually not surprising. Since the $f\leftrightarrow F$ correspondence is one-to-one up to pure gauge, and since pure gauge goes into pure gauge under \CG, it is clear that we will have a bona fide transformation of $f$ in the sense that any ambiguity in lifting $f$ to the cone will drop out. But the Euclidean fields which transform into themselves under the conformal group are exactly the primary fields. The only question is the interpretation of the $\Delta$ parameter, and an explicit analysis shows that it has the meaning of the scaling dimension.
 
\underline{To summarize}: instead of working with primary tensor fields in the physical space, we can do the computations with tensor fields in \ES, where \CG\ invariance is manifest, and project the result to \PS\ using (\ref{proj}).  Conformal invariance of the final result will be automatic.

\subsection{Correlators: Simplest Examples}

The embedding formalism provides a shortcut to solving constraints imposed by conformal symmetry on the form of CFT correlators. Consider e.g.\ the correlator of three primary scalars $\langle \phi_1(x_1)\phi_2(x_2)\phi_3(x_3)\rangle$ of dimensions $\Delta_i$. It can be obtained by projecting the embedding correlator
\beq 
\langle \Phi_1(P_1)\Phi_2(P_2)\Phi_3(P_3)\rangle=
\frac{const}{(P_{12})^{\frac{\Delta_1+\Delta_2-\Delta_3}{2}}
(P_{23})^{\frac{\Delta_2+\Delta_3-\Delta_1}{2}}
(P_{31})^{\frac{\Delta_3+\Delta_1-\Delta_2}{2}}}\,,
\label{scalar3pt}
\eeq
where we define
\beq
P_{ij}\equiv -2P_i\cdot P_j \,.
\eeq
It's easy to see that the written form of the correlator is the only one consistent with the \CG\ invariance and the degree $-\Delta_i$ homogeneity of each $\Phi_i(P_i)$. For scalars, projection to the physical space amounts to $P_i\to P_{x_i}$. Using the identity
\beq
-2P_{x_i}\cdot P_{x_j}=x^2_{ij} \qquad(x_{ij}\equiv x_i-x_j)\,,
\eeq
we obtain the well-known result \cite{3-pt}
\beq 
\langle \phi_1(x_1)\phi_2(x_2)\phi_3(x_3)\rangle=
\frac{const}{(x^2_{12})^{\frac{\Delta_1+\Delta_2-\Delta_3}{2}}
(x^2_{23})^{\frac{\Delta_2+\Delta_3-\Delta_1}{2}}
(x^2_{31})^{\frac{\Delta_3+\Delta_1-\Delta_2}{2}}}\,.
\eeq
Clearly, the embedding space derivation is more economical than the physical space one, which relies on the transformation properties of the primary scalars under   inversion.

As a second example, consider the two-point function $\langle v_a(x_1) v_b(x_2)\rangle$ of a dimension $\Delta$ primary vector, described in the embedding formalism by the correlator
\beq 
G_{AB}(P_1,P_2)\equiv \langle V_A(P_1) V_B(P_2)\rangle\,.
\eeq
$G_{AB}$ must be an \CG\ tensor satisfying the following properties:
\begin{eqnarray}
&G_{AB}(\lambda P_1,P_2) =G_{AB}(P_1,\lambda P_2)=  \lambda^{-\Delta}G_{AB}(P_1,P_2) \,,
\\
&P_1^A\,G_{AB}(P_1,P_2) =0\,, \ \ \ \ \ \ \ \ P_2^B\,G_{AB}( P_1,P_2) =0\,,
\label{conds}
\end{eqnarray}
following from the homogeneity and transversality conditions obeyed by $V_A(P)$. It is not difficult to convince oneself that the most general such tensor has the form
\beq
G_{AB}(P_1,P_2) =\frac{1}{(P_{12})^{\Delta}}\left[c_1 \tilde W_{AB}+c_2 \frac{P_{1A}P_{2B}}{P_1\cdot P_2}\right]\,,
\eeq
where
\beq
\tilde W_{AB} =   \eta_{AB} -\frac{P_{1B}P_{2A}}{P_1\cdot P_2}\,.
\label{W}
\eeq
(The reason for the tilde in $W$ will become clear shortly.) It remains to project to the physical space, using Eqs.~(\ref{proj}) and (\ref{expl}). The second term in $G_{AB}$ is pure gauge and projects to zero. A short computation shows that  $\tilde W_{AB}$ projects to
\beq
w_{ab}=\delta_{ab} - 2\, \frac{(x_{12})_a (x_{12})_b}{x_{12}^2}\,,
\eeq
and we get the well-known result
\beq
\langle v_a(x_1) v_b(x_2)\rangle ={c_1}{w_{ab}\over (x^2_{12})^{\Delta}}\,.
\label{spin1}
\eeq

The spin 2 case is analogous but with more indices. The embedding space two-point function is given by (up to pure gauge terms)\footnote{The same expression with all $\tilde W$'s replaced by $W$'s would work as well, differing only by pure gauge terms. We choose the given form to facilitate comparison with projector $\Pi'$ used in Eq.~\reef{Pi'} below.}
\beq
G_{A_1A_2,B_1B_2}(P_1,P_2)=\frac{const}{(P_{12})^{\Delta}}\left[\frac 12 \left(\tilde W_{A_1B_1} \tilde W_{A_2B_2}+\tilde W_{A_1B_2} \tilde W_{A_2B_1}\right) - \frac{1}{d}\, W_{A_1A_2} W_{B_1B_2}\right]\  ,
\label{EmbSpin2}
\eeq
where we introduced the symmetric tensor
\beq
W_{AB} =   \eta_{AB} -\frac{P_{1B}P_{2A}+P_{1A}P_{2B}}{P_1\cdot P_2}\,,
\label{Wno}
\eeq
differing from $\tilde W$ by a pure gauge term. Since both $W$ and $\tilde W$ are transverse, so is the above two-point function. To show that it is also traceless, notice that
\beq
 \eta^{A_1A_2} W_{A_1A_2} = d\,,
 \qquad  \eta^{A_1A_2}\tilde W_{A_1B_1} \tilde W_{A_2 B_2}=W_{B_1 B_2}\ .
 \label{trproj}
\eeq
Finally, the physical space two-point function is now obtained by projecting, which amounts to replacing  $W,\tilde W \to w$.

The generalization to higher $l$ is, in principle, straightforward. The two-point function can always be given by a symmetrized product of $\tilde W_{A_i B_j}$ with trace terms subtracted using $W_{A_i A_j}$. However, the computations become increasingly cumbersome due to the proliferation of indices, particularly if we wish to compute three-point and four-point functions. 
It would be nice to have a more compact formalism, which for example would allow not to keep track of the trace terms.
That this should be possible is intuitively clear, since these terms are not independent: they are fixed by the requirement of the overall tracelessness. In the next section we will describe such a formalism, which also has the advantage of being \emph{index-free}.

\section{Encoding Tensors by Polynomials}
\label{sec:tens}

To begin, we will introduce a technique which allows us to represent symmetric tensors by means of polynomials obtained by contracting the tensor with a reference vector. While the basic idea is very simple, it requires some effort to develop an efficient formalism fully taking into account the tracelessness and transversality conditions. The reader may prefer to read backwards starting from the example given in section \ref{sec:ex}. The less essential parts (proofs) are given in smaller font and can be  skipped on the first reading.

\subsection{Tensors in the Physical Space}
\label{sec:tens.phys.}

The basic idea is that any {symmetric} tensor can be encoded by a $d$-dimensional polynomial:
\beq
 f_{a_1\dots a_l} \ \text{symmetric}\leftrightarrow f(z)\equiv f_{a_1\ldots a_l} z^{a_1}\cdots z^{a_l} .
\label{code0}
\eeq
The correspondence is clearly one-to-one: expanding the polynomial we recover the tensor.
 
In CFT, spin $l$ primary fields are symmetric \emph{traceless} tensors, for which a more economical encoding is available. Such a tensor can be fully encoded by restricting the respective polynomial $f(z)$ to the submanifold $z^2=0$:\footnote{Assuming $z$ is complex.}
\beq
f_{a_1\ldots a_l} \ \text{symmetric traceless}\leftrightarrow f(z)|_{z^2=0} .
\label{code}
\eeq
This fact can be formulated more fully as follows. Let $f_{a_1\ldots a_l}$ be a symmetric traceless tensor, and  $\tilde{f}_{a_1\ldots a_l}$ be another symmetric  tensor such that the polynomials $\tilde{f}(z)$ and $f(z)$ differ only by terms vanishing on $z^2=0$:
\beq
f(z)=\tilde f(z)+O(z^2).
\label{agree}
\eeq
Then $f_{a_1\ldots a_l}$ can be recovered from $\tilde f(z)$ (or from $\tilde{f}_{a_1\ldots a_l}$, which is the same). 

Intuitively, this can be justified as follows.\footnote{A mathematician's proof that the correspondence (\ref{code}) is one-to-one goes as follows. First, observe that symmetric traceless tensors are mapped by (\ref{code0}) onto \emph{harmonic} polynomials.
Then, use the following theorem (see \cite{SW}, section 4.2): \emph{Any $d$-dimensional polynomial $p(z)$ can be uniquely split as $p(z)=p_0(z)+z^2\, p_1(z)$, with $p_0(z)$ harmonic.}}  
Consider the projector onto symmetric traceless tensors:
\beq
\pi_{a_1\ldots a_l,b_1\ldots b_l}=\delta_{a_1(b_1}\cdots \delta_{|a_l|b_l)}-\text{traces} \,.
\label{project}
\eeq
Eq.~(\ref{agree}) means that $f_{a_1\ldots a_l}$ and $\tilde f_{a_1\ldots a_l}$ can differ only by terms proportional to $\delta_{a_i a_j}$. All such terms will be subtracted away by the projector, and thus we will have:
\beq
 f_{a_1\ldots a_l}=\pi_{a_1\ldots a_l,b_1\ldots b_l}  \tilde f^{b_1\ldots b_l}\,.
 \label{recover}
 \eeq

\underline{To summarize the discussion so far}: we will present results for physical-space correlators in terms of polynomials, not in terms of tensors. Moreover, we can and will drop any polynomial terms explicitly proportional to $z^2$. This gives a polynomial which encodes the original symmetric traceless tensor in the sense of Eq.~(\ref{agree}). The dropped terms do not create any ambiguity, as the original tensor can be recovered via (\ref{recover}). 
\vspace{2em}

For small values of $l$, the projector appearing in (\ref{recover}) is easy to work out explicitly, e.g.
 \beq
\pi_{a_1 a_2,b_1 b_2}=\frac{1}{2}\left(\delta_{a_1b_1}\delta_{a_2b_2}+\delta_{a_1b_2}\delta_{a_2b_1} \right) -\frac{1}{d}\,\delta_{a_1a_2}\delta_{b_1b_2}\, .
\label{projspin2}
\eeq
The higher-spin projectors can be generated efficiently\footnote{An alternative is to use recursion relations, see e.g.\ \cite{Guth}.} by the differential operator of \cite{Todorov}\footnote{See \cite{Belitsky:2007jp} for a recent use of this operator in a similar context. It was also pointed out to us by Andrew Waldron that this operator appears in the context of `tractor calculus', where it is called the Thomas operator \cite{Waldron}.}
\beq
D_a=\left(h-1+z\cdot \frac{\partial}{\partial z} \right) \frac{\partial}{\partial z^a} -\frac{1}{2} z_a  
\frac{\partial^2\ }{\partial z \cdot \partial z}\,,
\label{Tod}
\eeq
where we defined the shorthand $ h \equiv d/2$. We then have
\beq
\pi_{a_1\ldots a_l,b_1\ldots b_l}={1 \over l!(h-1)_l}D_{a_1}\cdots D_{a_l} z_{b_1} \cdots z_{b_l}\,, 
\label{piTod}
\eeq
where $(a)_l=\Gamma(a+l)/\Gamma(a)$ is the Pochhammer symbol. It follows that $f_{a_1\ldots a_l}$ can be recovered from a $\tilde f(z)$ by differentiation:
\beq
f_{a_1\dots a_l}  = \frac{1}{l! (h-1)_l} D_{a_1}\cdots D_{a_l} \tilde f(z)\,.
\label{fTod}
\eeq

The $D_a$ operator is very convenient as it allows to perform operations on traceless symmetric tensors directly in terms of the polynomials that encode them. For example, consider two rank $l$ symmetric traceless tensors $f$ and $g$, encoded (in the sense of Eq.~(\ref{agree})) by $\tilde f(z)$ and $\tilde g(z)$. Then their full contraction can be found by evaluating
\beq
f_{a_1\dots a_l}g^{a_1\dots a_l}  = \frac{1}{l! (h-1)_l} \tilde f(D) \tilde g(z)\,.
\label{contr}
\eeq
If we need to free just one index but leave the rest contracted with $z$, this is done by evaluating
\beq
f_{a a_2\dots a_l}z^{a_2}\cdots z^{a_l}  = \frac{1}{l (h+l-2)} D_a \tilde f(z)+O(z^2)\,,
\label{one}
\eeq
and so on.

\vspace{1em}
{\proof We will just give a general idea of how these statements can be proven; see Appendix A of \cite{Todorov-book} for more details. It is crucial that $D_a$ is an `interior operator' on the cone, which means that it maps $O(z^2)$ functions to themselves:
\beq
h(z)=O(z^2)\ \Longrightarrow\ D_a h(z)=O(z^2)\,.
\label{int}
\eeq
In particular, we have
\beq
D_a \tilde f(z)=D_a f(z)+O(z^2) .
\eeq
Furthermore, tracelessness of $f$ implies that the polynomial $f(z)$ is harmonic. Thus the second term in $D_a$ does not contribute to $D_a f$, while the first term gives
\beq
D_a f(z)=\left(h-1+z\cdot \frac{\partial}{\partial z} \right) \frac{\partial}{\partial z^a} f(z)=(h+l-2) l\, f_{a a_2\dots a_l}z^{a_2}\cdots z^{a_l} , 
\eeq
where we used the fact that $z\cdot \frac{\partial}{\partial z}$ computes the homogeneity degree, $l-1$ in this case. This proves Eq.~(\ref{one}); the other properties can be shown analogously.}

\subsection{Tensors in the Embedding Space}
\label{sec:tens.emb.}

Next we will extend the above discussion to the embedding space.  We can similarly encode a general symmetric tensor in the embedding space by a ($d+2$)-dimensional polynomial 
\beq
F_{A_1\ldots A_l}(P) \ \text{symmetric}\leftrightarrow F(P,Z)\equiv F_{A_1\ldots A_l}(P) Z^{A_1}\ldots Z^{A_l} .
\label{Zcode0}
\eeq
This notation emphasizes that the tensors will in general depend on $P$. 

Now let us consider the following diagram relating embedding and physical tensors both with free indices and with encoding polynomials:
\beq
\xymatrix{
F_{A_1\ldots A_l}(P) \ar[r]^-{(\ref{Zcode0})} \ar[d]_-{(\ref{proj})} & F(P,Z)  \ar@{-->}[d]\\
f_{a_1\ldots a_l}(x)  \ar[r]^-{(\ref{code0})}  &  f(x,z)
} 
\eeq
The dashed line denotes that there is a relation between the encoding polynomial of an embedding tensor and its projection to the physical space. Using the explicit form of $\partial P/\partial x$ given in Eq.~(\ref{expl}), 
this relation takes the form
\beq
f(x,z)=F(P_x,Z_{z,x}),
\label{link}
\eeq
where $Z_{z,x}\equiv(0,2x\cdot z, z)$ and has the properties
\beq
Z_{z,x}\cdot P_x=0\,,\qquad Z_{z,x}^2=z^2\,.
\label{Zprop}
\eeq

\vspace{0.5em}
Let us now specialize to tensors which are symmetric, traceless, and transverse (STT). For such tensors, we can restrict the polynomial to the subset of $Z$'s satisfying $Z^2=0$ and $Z\cdot P=0$:
 \beq
F_{A_1\ldots A_l}(P)\quad \text{STT}\qquad\leftrightarrow\qquad F(P,Z)|_{Z^2=0,Z\cdot P=0} \,.
\label{Zcode}
\eeq
More precisely, we mean the following. Let $F_{A_1\ldots A_l}(P)$ be STT and $\tilde F_{A_1\ldots A_l}(P)$ be any tensor whose polynomial happens to agree with $F(P,Z)$ modulo terms proportional to $Z^2$ and $Z\cdot P$:
\beq
F(P,Z)=\tilde F(P,Z)+O(Z^2,Z\cdot P) .
\label{Zagree}
\eeq
Then $F_{A_1\ldots A_l}(P)$ can be recovered from $\tilde F_{A_1\ldots A_l}(P)$ up to pure gauge terms.

\vspace{1em}
{\proof Indeed, as discussed in section \ref{sec:emb}, the tensor $F$ can be recovered up to pure gauge from its symmetric traceless projection $f$. Thus it is enough to show that $f$ can be determined from $\tilde f$, the projection of $\tilde F$. To see the latter, let us project Eq.~(\ref{Zagree}) to the physical space.  Using the rule (\ref{link}) and the properties (\ref{Zprop}), we obtain
\beq
f(x,z)=\tilde f(x,z)+O(z^2),
\label{ftildef}
\eeq
so $f$ can indeed be recovered from $\tilde f$ by one of the methods from section \ref{sec:tens.phys.}.}
\vspace{1em}

Since it will prove useful in future applications, let us give a more explicit way to recover an STT tensor $F_{A_1 \ldots A_l}(P)$ from $\tilde F_{A_1 \ldots A_l}(P)$ in the case that $\tilde F$ is transverse (but not necessarily traceless).  In this case the projection takes the form
\beq
 F_{A_1\ldots A_l}=\Pi_{A_1\ldots A_l,B_1\ldots B_l}  \tilde F^{B_1\ldots B_l}\, \label{recoverZ} ,
 \eeq
 where the projector $\Pi$ is obtained from the projector $\pi$ in Eq.~\reef{recover} by replacing 
 \beq
 \delta_{a_ia_j}\to W_{A_iA_j}\equiv \eta_{A_iA_j}-{P_{A_i}\bar P_{A_j}+P_{A_j}\bar P_{A_i} \over P\cdot \bar P},\quad  \delta_{b_ib_j}\to \eta_{B_iB_j},\quad  \delta_{a_ib_j}\to \eta_{A_iB_j}.
 \label{repl1}
 \eeq
Here $\bar P$ is as in Eq.~\reef{id1}. The rule may look strange, since the projector $\pi$ subtracts traces in $d$ dimensions, while $\Pi$ must do this in $d+2$ dimension. This connection between $\pi$ and $\Pi$ has to do with the assumed transversality of $\tilde F$.

\vspace{1em}
{\proof To prove that the above formula works, notice first of all that the tensor $F$ as defined differs from $\tilde F$ only by terms which are proportional to $\eta_{A_iA_j}$ or $P_{A_i}$. Upon contraction with $Z$, this gives terms of $O(Z^2,Z\cdot P)$, consistent with Eq.~\reef{Zagree}. It remains to show that $F$ is transverse and traceless. To this end, consider a different projector $\Pi'$ obtained from $\pi$ by a list of replacements which contains some extra terms compared to \reef{repl1}:
\begin{gather}
 \delta_{a_ia_j}\to W_{A_iA_j},\quad  \delta_{b_ib_j}\to W_{B_iB_j},
\quad  \delta_{a_ib_j}\to \tilde W_{A_iB_j}\equiv\eta_{A_i B_j}-{\bar P_{A_i}P_{B_j} \over P\cdot \bar P} .
\end{gather}
However, all the extra terms are proportional to $P_{B_i}$, and will vanish when contracted with $\tilde F$ under the assumption that it is transverse. For this reason we have an equivalent representation for $F$ as
\beq
F=\Pi' \tilde F .
\label{Pi'}
\eeq
In this form transversality and tracelessness are pretty easy to see. They just follow from the transversality of $W$ and $\tilde W$, and from the relations~\reef{trproj} that we already used to show that the spin-2 two-point function \reef{EmbSpin2} was transverse and traceless.  Indeed, as the reader may have noticed, that two-point function had precisely the structure of the traceless projector in $d$ dimensions, Eq.~\reef{projspin2}.
\vspace{1em}}

In this paper, we will be primarily dealing with tensors which are made from metrics and from components of \ES\ vectors, such as in Eq.~(\ref{EmbSpin2}). For such tensors, the canonical rule to get the encoding polynomial $\tilde F(P,Z)$ in Eq.~(\ref{Zagree}) is to simply drop all terms in $F(P,Z)$ which are proportional to $Z^2$ and $Z\cdot P$. This rule is also very convenient because it preserves the transversality condition, and even makes it stronger, in a sense that we now discuss.

In general, a transverse tensor $F_{A_1\ldots A_l}$ may contain terms which are pure gauge, and the condition $P\cdot F=0$ is only valid modulo $P^2$ terms, vanishing on the cone. We will call a tensor \emph{identically transverse} if this condition happens to be satisfied identically, without using $P^2=0$. For example, the tensor $\tilde W$ from Eq.~\reef{W} 
is identically transverse with respect to $P_1^A$ and $P_2^B$, while $W$ from Eq.~\reef{Wno} is not.  
Notice that $\tilde W$ can be obtained from $W$ by dropping the pure gauge term. This is in fact a partial case of the following more general rule: 

Take any tensor $F_{A_1\ldots A_l}(P)$ which is 
\begin{enumerate}
\item Transverse modulo $P^2$ terms. 
\item Made out of metrics and components of $P$, as well as of components of one or more vectors $Q\ne P$. 
\end{enumerate}
Drop any terms in the tensor which are proportional to $P^2$, $\eta_{A_iA_j}$, or $P_{A_i}$. The resulting tensor $\tilde F_{A_1\ldots A_l}(P)$ will be identically transverse.
 
 \vspace{1em}
{\proof To prove this, let us write
 $F=\tilde F+ \hat{F}$, where $\hat{F}$ contains all terms which are to be dropped. Then $P\cdot \tilde F$ will contain terms proportional to $Q_{A_i}$, with coefficients which are scalar functions of $(P\cdot Q)$ and $(Q\cdot Q')$ (if there are several $Q$'s). On the other hand, $P\cdot \hat F$ will contain terms proportional to $P_{A_i}$ and/or $P^2$.
There cannot be cancellation between these two groups of terms, and if $P\cdot F$ is to vanish on $P^2=0$, $P\cdot \tilde F$ must vanish identically.}
\vspace{1em}

 Going back to the encoding polynomials, the transversality condition takes the form
 \beq
P\cdot \frac{\partial}{\partial Z} \ F(P,Z)= 0 \,,
\label{trans-der}
\eeq
or equivalently
\beq
F(P,Z+\alpha P)= F(P,Z)  \qquad  (\forall\alpha)\,.
\label{trans}
\eeq
These conditions are satisfied modulo $P^2$ in general, and identically if the tensor is identically transverse. Translating the above discussion, the identically transverse polynomial $\tilde F(P,Z)$ is obtained from $F(P,Z)$ by dropping all terms proportional to $Z^2$ and $Z\cdot P$.  This is precisely the `canonical rule' introduced above.

The above discussion will prove very useful below, because the identically transverse polynomials are easy to characterize. It is not difficult to convince oneself that the following rule is true: a polynomial $\tilde F(P,Z)$ is identically transverse if and only if the variable $Z_A$ appears in it only via the tensor:
\beq
C_{AB} \equiv Z_A P_B-Z_B P_A\,.
\label{CAB}
\eeq

To conclude this section, let us show how to compute tensor contractions using the embedding space. The problem is formulated as follows. We want to contract two symmetric traceless tensors $f_{a_1\dots a_l}(x)$ and $g_{a_1\dots a_l}(x)$. It is assumed that these tensors are projections of the embedding space STT tensors $F_{A_1\dots A_l}(P)$ and $G_{A_1\dots A_l}(P)$. The latter tensors will typically not be given in components, but in terms of their encoding polynomials $\tilde F(P,Z)$ and $\tilde G(P,Z)$ (in the sense of Eq.~(\ref{Zagree})). Finally, we will assume that these polynomials are transverse in the sense of Eq.~(\ref{trans}).\footnote{Although not essential here, in applications they will often be even identically transverse.} We then have the formula (cf. Eq.~(\ref{fTod})):
\beq
f_{a_1\dots a_l}(x)g^{a_1\dots a_l}(x)  = \frac{1}{l! (h-1)_l} \tilde F(P_x,D) \tilde G(P_x,Z)\,,
\label{Zcontr}
\eeq
where
\beq
D_A=\left(h-1+Z\cdot \frac{\partial}{\partial Z} \right) \frac{\partial}{\partial Z^A} -\frac{1}{2} Z_A  
\frac{\partial^2\ }{\partial Z \cdot \partial Z} 
\label{EmbTod}
\eeq
is the same differential operator as $D_a$ made to act in the $(d+2)$-dimensional space. We stress that $h=d/2$ here as in Eq.~(\ref{Tod}).

\vspace{1em}
{\proof Let us give a quick proof. Using the notation of section \ref{sec:tens.phys.}, we have
\beq
f_{a_1\dots a_l} g^{a_1\dots a_l}= \tilde f_{a_1\dots a_l} \pi^{a_1\dots a_l,b_1\dots b_l}\tilde g_{b_1\dots b_l} =
 \tilde F_{A_1\dots A_l} Q^{A_1\dots A_l,B_1\dots B_l}\tilde G_{B_1\dots B_l}\,, 
\eeq
where $\tilde f$ and $\tilde g$ are the projections of $\tilde F$ and $\tilde G$ to the physical space, and $Q$ is given by 
\beq
 Q^{A_1\dots A_l,B_1\dots B_l} =\pi^{a_1\dots a_l,b_1\dots b_l} \frac{\partial P^{A_1}}{\partial x^{a_1}}\cdots  \frac{\partial P^{A_l}}{\partial x^{a_l}}  
  \frac{\partial P^{B_1}}{\partial x^{b_1}}\cdots   \frac{\partial P^{B_l}}{\partial x^{b_l}}\,.
  \label{Pi}
\eeq
Remember that the projector $\pi$ is made out of $d$-dimensional metric tensors. This equation then means that the projector $Q$ can be obtained from $\pi$ by replacing 
each metric $\delta^{ab}$ by the effective metric $K^{AB}$ defined in Eq.~(\ref{id1}) (unlike in the definition of $\Pi$ above, here the replacement rule is the same whether the indices are of $a$ or $b$ type). For transverse tensors $\tilde F$ or $\tilde G$ we can replace $K^{AB}$ by $\eta^{AB}$ because the extra terms vanish identically. A moment's thought shows that this reduces (\ref{Zcontr}) to (\ref{contr}).}

\subsection{Example}
\label{sec:ex}
Let us now demonstrate the above formal discussion on a concrete example: the spin 2 embedding space two-point function (\ref{EmbSpin2}). Since it's a double tensor, we assign to it a polynomial of two vectors $Z_1$ and $Z_2$:
\beq
G(P_1,P_2;Z_1,Z_2) = Z_1^{A_1}Z_1^{A_2}Z_2^{B_1}Z_2^{B_2} \,
G_{A_1A_2,B_1B_2}(P_1,P_2)\,.
\eeq
We then have the following basic contractions:
\begin{gather}
Z_1^A Z_2^B \tilde W_{AB}=(Z_1\cdot Z_2)-{(Z_1\cdot P_2)(Z_2\cdot P_1) \over P_1\cdot P_2} \,, \\
Z_1^A Z_1^{A'} W_{AA'}=O(Z_1^2,Z_1\cdot P_1)\,,\qquad Z_2^B Z_2^{B'} W_{BB'}=O(Z_2^2,Z_2\cdot P_2)\,.
\end{gather}
It follows that 
\beq
\tilde G(P_1,P_2;Z_1,Z_2) =const \,\frac{\big((Z_1\cdot Z_2)(P_1\cdot P_2)-(P_1\cdot Z_2)(P_2\cdot Z_1)\big)^2}{(P_{12})^{\Delta+2}}\,,
\label{l=2emb}
\eeq
where we applied the canonical rule of dropping the $O(Z_i^2,Z_i\cdot P_i)$ terms to get the encoding polynomial. Notice that $\tilde G$ is identically transverse, as it should be according to the discussion in section \ref{sec:tens.emb.}. This is already a pretty compact expression; the advantage of not having to deal with indices is starting to show.  

What about the two-point function in physical space? We will write it as a polynomial contracted with $z_1$ and $z_2$. This polynomial is obtained by making the substitutions 
$P_i\to P_{x_i}$, $Z_i\to Z_{z_i,x_i}$ in $\tilde G$. Evaluating the scalar products
\begin{gather}
Z_1\cdot Z_2\to z_1\cdot z_2\,,\qquad P_1\cdot P_2\to-\frac 12  \,x_{12}^2 \,,\\
P_1\cdot Z_2\to z_2\cdot x_{12}\,,\qquad P_2\cdot Z_1\to - z_1 \cdot x_{12} \,,
\end{gather}
we find
\beq
g(x_1,x_2;z_1,z_2) =const\, \frac{\bigl((z_1\cdot x_{12})(z_2\cdot x_{12})-\frac 12 x_{12}^2(z_1\cdot z_2)\bigr)^2}{(x^2_{12})^{\Delta+2}}\,,
\eeq
up to $O(z_i^2)$ terms (see Eq.~(\ref{ftildef})). In the index-free approach that we are advocating here, this expression is \emph{the} final answer. The indexed version can be extracted if necessary by acting with $D_a$ operators on the encoding polynomial, or in a more pedestrian way, by expanding in $z^a_i$ and acting on the coefficient tensor with the projector $\pi$. But in this paper we will not do this.

\section{Correlation Functions of Spin $l$ Primaries}
\label{sec:corr}
Unitary irreducible representations of the conformal group $SO(d+1,1)$ are labeled by a conformal dimension $\Delta$ and an irreducible representation of $SO(d)$. In this paper, we focus on totally symmetric traceless tensors of $SO(d)$.
These are the spin $l$ primaries, which we will label by $\chi\equiv[l,\Delta]$. In this section, we discuss constraints imposed by  conformal symmetry on the coordinate dependence of their correlators. The additional constraints appearing for conserved tensors will be discussed in the next section. 

\subsection{Two-Point Functions}
\label{2pt}

Consider the two-point function of a spin $l$ primary in the embedding space:
\beq
G_{A_1\ldots A_l,B_1\ldots B_l}(P_1,P_2)\,.
\eeq
Following the technique from the previous section, we will encode it by a function 
\beq
G_\chi(P_1,P_2;Z_1,Z_2) = Z_1^{A_1}\cdots Z_1^{A_l}Z_2^{B_1}\cdots Z_2^{B_2} \,
G_{A_1\ldots A_l,B_1\ldots B_l}(P_1,P_2)\,.
\eeq
We have the following three conditions:
\begin{gather}
G_\chi(\lambda_1 P_1, \lambda_2 P_2;  Z_1,Z_2) =(\lambda_1 \lambda_2 )^{-\Delta}G_\chi(P_1, P_2; Z_1 , Z_2 )\,,\\
G_\chi(P_1, P_2;  \beta_1Z_1,\beta_2Z_2) =(\beta_1 \beta_2 )^{l}G_\chi(P_1, P_2; Z_1 , Z_2 )\,,\\
G_\chi(P_1, P_2;  Z_1+\alpha_1P_1,Z_2+\alpha_2 P_2) =G_\chi(P_1, P_2; Z_1 , Z_2 )\,. 
\label{chi-trans}
\end{gather}
The first condition follows from the fact that the embedding space fields are homogeneous of degree $-\Delta$. The second one is a fancy way of saying that $G_\chi$ is a degree $l$ polynomial in $Z_1$ and $Z_2$. The final condition encodes the transversality of the embedding space tensors; it must be satisfied modulo $O(P^2)$ terms.

As discussed in section \ref{sec:tens.emb.}, we may drop all the terms in $G_\chi$ proportional to $Z_i^2$ and $Z_i\cdot P_i$. The resulting function $\tilde G_\chi$ will be identically transverse, in the sense that it will satisfy Eq.\ (\ref{chi-trans}) identically, and not just modulo $O(P^2)$. The general recipe for constructing such functions says that they must be built out of the $C_{AB}$-type tensors from Eq.\ (\ref{CAB}):
\beq
C_{iAB}=Z_{iA} P_{iB}-Z_{iB} P_{iA}\qquad(i=1,2)\,.
\eeq
Now contracting $C_{i}$ with itself gives terms of the kind that we dropped, and so the only possibility is to start contracting the indices of $C_{1}$ and $C_{2}$. Full contraction gives the  building block
\beq
H_{12} \equiv - C_{1}\cdot C_{2}=- 2\big[(Z_1\cdot Z_2)(P_1\cdot P_2)-(P_1\cdot Z_2)(P_2\cdot Z_1)\big]\,,
\label{H12}
\eeq
of weight one in both $Z_1$ and $Z_2$.
More generally, one could try taking the trace of a string of several alternating $C_1$'s and $C_2$'s. However, one can check that
\beq
(C_{1}C_{2}C_1)_{AB}=-\frac 12 \left(C_{1}\cdot C_{2}\right) C_{1AB}\,.
\eeq
For this reason, such iterated contractions reduce to powers of $C_{1}\cdot C_{2}$.
We conclude that the most general solution is a function of $C_{1}\cdot C_{2}$. The spin of the operators fixes the weight in the $Z$'s, so we obtain that
(cf. Eq. (\ref{l=2emb}))
\beq
\tilde G_\chi(P_1,Z_1;P_2,Z_2) =const \,\frac{H_{12}^l}{(P_{12})^{\Delta+l}}\,.
\eeq
Thus we recover the well-known unique two-point function of spin $l$ primaries \cite{FGG-book}.

\subsection{Three-Point Functions}\label{3pt}

The scalar three-point function was already given in Eq.~(\ref{scalar3pt}).
In this section we will discuss the arbitrary spin case using the embedding formalism.
It is well known that such three-point functions can be written as a linear combination of a finite number of 
conformally invariant building blocks \cite{Mack, Zaikov,OP,MP,Giombi}.
Here, we present the explicit form of these building blocks in the embedding formalism.

\subsubsection{Scalar-Scalar-Spin $l$}

Let us start with the scalar-scalar-spin $l$ case. The scalar operators of dimensions $\Delta_1$ and $\Delta_2$ are placed at points $P_1$ and $P_2$. The third operator, a symmetric traceless tensor of spin $l$ and dimension $\Delta_3$, is placed at $P_3$.
In this case, the correlator 
is completely fixed by conformal invariance. We have ($l_3=l$)
\begin{gather}
\tilde G_{\chi_1,\chi_2,\chi_3} (P_1,P_2,P_3;Z_3) = 
const\,\frac{\big((Z_3\cdot P_1)(P_2\cdot P_3)-(Z_3\cdot P_2)(P_1\cdot P_3)\big)^{l}}
{(P_{12})^{\frac{\Delta_1+\Delta_2-\Delta_3+l}{2}}
(P_{23})^{\frac{\Delta_2+\Delta_3-\Delta_1+l}{2}}
(P_{31})^{\frac{\Delta_3+\Delta_1-\Delta_2+l}{2}}}\,.
\label{ScalarScalarSpinl}
\end{gather}
Here we are using the same notation as in the two-point function case. The polynomial $\tilde G_{\chi_1,\chi_2,\chi_3}$ is obtained from the correlator polynomial $G_{\chi_1,\chi_2,\chi_3}$ by dropping all terms proportional to $Z_3^2$ and $Z_3\cdot P_3$. This polynomial must be identically transverse, and so it must be constructed out of the tensor $C_{3AB}$. The only possibility is to contract this tensor with $P_1$ and $P_2$, which gives the structure
\beq
V_{3,12} \equiv \frac{P_1\cdot C_3 \cdot P_2}{P_1\cdot P_2} =\frac{ (Z_3\cdot P_1)(P_2\cdot P_3)-(Z_3\cdot P_2)(P_1\cdot P_3)}{P_1\cdot P_2}
\label{V312}
\eeq
used in (\ref{ScalarScalarSpinl}). The exponents are then fixed by the homogeneity requirements.

\subsubsection{General Spins $l_1,l_2$  and $l_3$}

We now proceed to the general case of the three-point function of symmetric
traceless operators of spins $l_i$. We will write it as
\beq
\tilde G_{\chi_1,\chi_2,\chi_3} (\{ P_i;Z_i\}) = 
\frac{Q_{\chi_1,\chi_2,\chi_3} (\{ P_i;Z_i\})}
{(P_{12})^{\frac{\tau_1+\tau_2-\tau_3}{2}}
(P_{23})^{\frac{\tau_2+\tau_3-\tau_1}{2}}
(P_{31})^{\frac{\tau_3+\tau_1-\tau_2}{2}}}\,,
\label{general3pt}
\eeq
where $\tau_i=\Delta_i+ l_i$. 
The numerator $Q_{\chi_1,\chi_2,\chi_3} (\{ P_i;Z_i\})$ is an identically transverse polynomial of degree $l_i$ in each $Z_i$, with coefficients which depend on $P_i$. With the above normalization, $Q$ is also homogeneous of degree $l_i$ in each $P_i$. Thus,
\beq
Q_{\chi_1,\chi_2,\chi_3}(\{\lambda_i P_i; \alpha_i Z_i+\beta_i P_i\})=
 Q_{\chi_1,\chi_2,\chi_3}(\{P_i;Z_i\})\  
\prod_i \left(\lambda_i  \alpha_i \right)^{l_i} \,.
\label{Qconditions3pt}
\eeq

According to the general characterization of transverse polynomials, $Q$ must be built by contracting the tensors $C_{iAB}$ among themselves and with vectors $P_i$. Not all contractions are useful, since $C_i\cdot C_i$, $C_i\cdot P_i$, $C_i\cdot Z_i$ give terms proportional to $Z_i^2$ and $Z_i\cdot P_i$ which are to be dropped. 

Examples of nontrivial building blocks are given by contractions using different points, for instance $C_{1}\cdot C_2$ in (\ref{H12}) and 
$P_1\cdot C_{3}\cdot P_2$  in (\ref{V312}). It is then clear that three-point functions can be constructed from the basic building blocks
\begin{align}
V_{i,jk} &\equiv\frac{P_j\cdot C_i \cdot P_k}{P_j\cdot P_k} =\frac{(Z_i \cdot P_j )(P_i\cdot P_k )- (Z_i \cdot P_k )( P_i\cdot P_j)}{(P_j \cdot P_k)}\,,
\label{V_i,jk} \\
H_{ij} & \equiv - C_{i}\cdot C_{j} =- 2\big[ (Z_i \cdot Z_j )( P_i\cdot P_j) - (Z_i \cdot P_j )( Z_j\cdot P_i)\big]\,,
\label{H_ij}
\end{align}
which are transverse. They also satisfy the scaling conditions (\ref{Qconditions3pt}) with
$l_i=1$, $l_j=l_k=0$ for $V_{i,jk}$; $l_i=l_j=1$, $l_k=0$ for $H_{ij}$.

However, not all $V_{i,jk}$ and $H_{ij}$ are linearly independent 
due to $V_{i,jk}=-V_{i,kj}$ and $H_{ij}=H_{ji}$. Hence there are three linearly independent $V$'s and three linearly independent $H$'s. Explicitly we
will use the following basic structures 
\begin{equation}
V_1 \equiv V_{1,23} \,, \quad V_2 \equiv V_{2,31} \,, \quad V_3 \equiv V_{3,12} \,, \quad H_{12}\,, \quad   H_{13}\,, \quad  H_{23}\,.
\end{equation}

In principle, one could imagine more complicated contractions involving several $C_i$'s. However, it turns out that they will not produce any new structure. Namely, \emph{any identically transverse polynomial Q can be written as a function of $V_i$ and $H_{ij}$ only (with $P$-dependent coefficients)}. For the simplest examples, like $\text{Tr}[C_1C_2C_3]$, this can be checked by an explicit computation. A general proof can be given as follows:

\vspace{1em}
{\proof First, take the special case when $Q$ is identically transverse and depends only on $Z_i\cdot P_j$ but not on $Z_i\cdot Z_j$. It is easy to convince oneself that such a $Q$ must be a function of $V_i$. In the general case, let us first rewrite $Q$ by expressing all $Z_i\cdot Z_j$ products via $H_{ij}$ from Eq.\ (\ref{H_ij}). This of course generates new terms, which are however all proportional to $Z_i\cdot P_j$. This shows that $Q$ can be expressed as a polynomial in $H_{ij}$ with coefficients which are functions of $Z_i\cdot P_j$. Moreover, from the way we arrived at this representation, it's clear that it is unique. In this representation, the transversality of $Q$ implies the transversality of all the coefficients (since $H_{ij}$'s are transverse by themselves). According to the special case treated first, these coefficients can be written as functions of $V_i$.}
\vspace{1em}

The conclusion of the above discussion is that the general solution for $Q_{\chi_1,\chi_2,\chi_3}$ can be written as a linear combination of 
\beq
\prod_i V_i^{m_i} \ \prod_{i<j} H_{ij}^{n_{ij}}\,. \label{Embterm}
\eeq
Since $Q$ must have degree $l_i$ in each $Z_i$, the exponents must satisfy the three constraints
\beq
m_i +\sum_{j\neq i} n_{ij} =l_i\,.  \label{expconstraints}
\eeq
These equations imply as well that $Q$ has degree $l_i$ in each $P_i$, as it should.  Notice that with three $P_i$'s at our disposal, we cannot construct any nontrivial functions of $P_i$ of zero homogeneity (with four $P_i$'s this would be possible; see the four-point function case below). This means that there is no further ambiguity in the coordinate dependence of $Q$.

\begin{figure}[htbp]
\begin{center}
\includegraphics[scale=0.4]{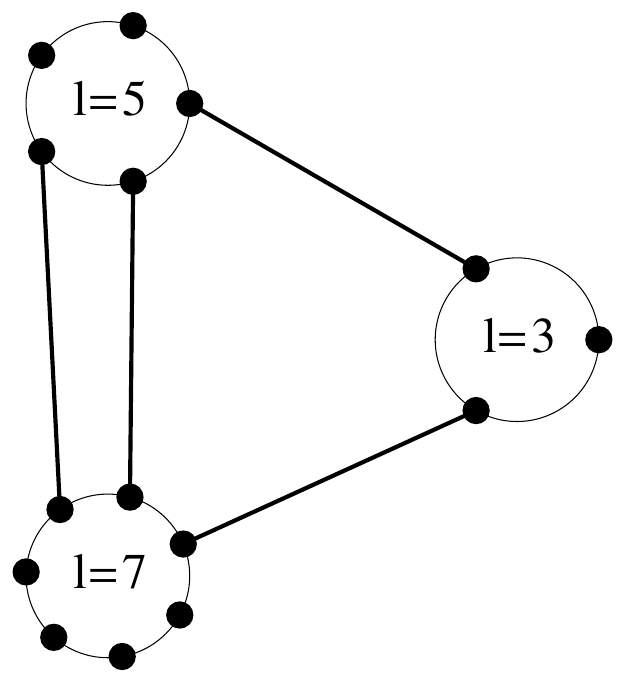}
\caption{Schematic representation of one of the tensor structures appearing in the (spin 5)-(spin 3)-(spin 7) three-point function. $V_i$'s are represented as disconnected dots at the vertices and $H_{ij}$'s as lines joining the vertices.}
\label{Diagram3pt}
\end{center}
\end{figure}

Eq.\ (\ref{Embterm}) implies that for general spins $l_i$ there will be several inequivalent three-point function structures compatible with the conformal symmetry. Their number is equal to the number of non-negative integer points $(n_{12},n_{13},n_{23})$ in the three dimensional polyhedron defined by the conditions
\beq
n_{12}+n_{13} \le l_1\,, \ \ \ \ \ \ \ \ \ \ \ 
n_{12}+n_{23} \le l_2\,, \ \ \ \ \ \ \ \ \ \ \ 
n_{13}+n_{23} \le l_3\,.
\eeq
Counting these points, it is possible to write the number of inequivalent structures in closed form:
\beq
N(l_1,l_2,l_3)=\frac{(l_1+1)(l_1+2)(3l_2-l_1+3)}{6} - \frac{p(p+2)(2p+5)}{24} - \frac{1-(-1)^p}{16} \,,
\label{N3pt}
\eeq
where we have ordered the spins $l_1\le l_2\le l_3$ and defined $p\equiv {\rm max} (0,l_1+l_2-l_3)$.

\vspace{1em}
{\bf Note added:} When this paper was being finalized, Ref.~\cite{Paulos:2011ie} appeared which among other things points out that  conformal structures corresponding to operators with spin can be constructed from a smaller set of elementary structures.  Our structures $V_{i,jk}$ and $H_{ij}$ are index-free equivalents of the structures $X_{ij}^{M_k}$ and $I^{M_iM_j}$ appearing in~\cite{Paulos:2011ie}. We believe that our index-free formalism is cleaner and more versatile, especially when various degeneracies among basic structures need to be taken into account, as in several situations discussed below, and also when considering traceless tensors.

\subsubsection{Parity Odd Three-Point Functions}
\label{parityodd}

So far we have implicitly assumed that the correlators are parity invariant.
If this is not the case, then there are additional structures in the three-point function.
More precisely, we can use the \CG-invariant $\epsilon$-tensor to construct new building
blocks for the three-point function. Since the product of two $\epsilon$-tensors can be written
in terms of metrics, it is enough to use the $\epsilon$-tensor once. 
The number of invariant structures that can be built from one $\epsilon$-tensor
and the vectors $P_i$ and $Z_i$ depends on the dimension $d$. For $d>4$ it is not possible to form a scalar from these ingredients. This implies that all conformally
invariant  three-point functions of spin $l_i$ symmetric traceless operators 
in $d>4$ are necessarily parity invariant.\footnote{Sometimes the correlators containing $\epsilon$-tensors are called parity violating in the literature, which is poor terminology. The theory may be perfectly parity preserving even though some correlators are parity odd, provided that the fields themselves are assigned negative parity. 
A notable exception is the stress tensor, which must be assigned positive parity by its very meaning as the generator of spacetime transformations, and also more formally since the correlator $\langle T T T\rangle$ necessarily contains a parity even term (due to the Ward identity)~\cite{OP}. In this case, any admixture of a parity odd structure~\cite{MP,Giombi} would imply parity violation.}

For $d=4$, there is a unique invariant
\beq
\epsilon(Z_1,Z_2,Z_3,P_1,P_2,P_3) \,,
\label{epsilon(d=4)}
\eeq
where by $\epsilon(\cdots)$ we mean the contraction of the $(d+2)$-dimensional  $\epsilon$-tensor
with all the arguments.
Thus, the number of parity odd  structures of $(l_1,l_2,l_3)$  three point  functions is equal
to the number of parity even structures of $(l_1-1,l_2-1,l_3-1)$  three point  functions, since 
(\ref{epsilon(d=4)}) involves a single power in each $Z_i$.

For $d=3$, there are 3 invariants
\beq
\epsilon(Z_i,Z_j ,P_1,P_2,P_3) \,.
\eeq
Notice that $\epsilon(Z_1,Z_2 ,Z_3,P_1,P_2)$ is not invariant under
$Z_3 \to Z_3 +\beta P_3$ and therefore is excluded.
In fact, in 3 dimensions not all conformally invariant building blocks are independent. We treat this special case separately in section  \ref{3D3pt}.

\subsubsection{Relation to Leading OPE Coefficient} \label{OPvsEmb}
 
Mack \cite{Mack} and Osborn and Petkou \cite{OP} give a prescription to uplift the leading OPE coefficient into a conformally invariant three point 
function. Here we wish to make direct contact with this work, starting from the embedding formalism. 

Let us rewrite Eq.~(2.31)  of  \cite{OP} as follows:
\beq
\mathcal{O}_1(x ,z_1)\mathcal{O}_2(0,z_2) \sim \mathcal{O}_3(0,\del_{z_3})\, t(x,z_1,z_2,z_3)\,
x ^{-(\Delta_1+\Delta_2-\Delta_3+\sum l_i)} ,
\eeq
where 
$x^\alpha$ stands for $(x^2)^{\frac{\alpha}{2}}$, and
\beq
\mathcal{O}(x ,z)=z^{\mu_1}\cdots z^{\mu_l} \mathcal{O}_{\mu_1\dots\mu_l}(x) \,.
\eeq 
The choice of a rotationally invariant tensor structure for the leading OPE coefficient is the choice of rotationally invariant polynomial $t$ such that
\beq
t(\lambda x,\lambda_1z_1,\lambda_2z_2,\lambda_3z_3)=
t(x,z_1,z_2,z_3)
\prod_{i=1}^3 (\lambda \lambda_i)^{l_i}\,.
\eeq
Equation (2.36) of \cite{OP} then becomes 
\beq
\langle \mathcal{O}_1(x_1 ,z_1)\mathcal{O}_2(x_2,z_2) \mathcal{O}_3(x_3 ,z_3)\rangle= \frac{t\left(X_{12},\tilde{z}_1,\tilde{z}_2 ,z_3\right)
}{x_{13}^{2\Delta_1} x_{23}^{2\Delta_2}
X_{12}^{\Delta_1+\Delta_2-\Delta_3+\sum l_i}}\,,
\eeq
where
\beq
X_{12}=\frac{x_{13}}{x_{13}^2}-\frac{x_{23}}{x_{23}^2}\ ,\ \ \ \ \ 
\tilde{z}_1=R(x_{13})z_1\ ,\ \ \ \ \ 
\tilde{z}_2=R(x_{23})z_2\ ,
\eeq
where $R(x)$ is a linear transformation acting on $z_i$ as
\beq
R(x)_{\mu\nu}=\delta_{\mu\nu}-\frac{2x_\mu x_\nu}{x^2}\,.
\eeq
Using $X_{12}^2=x_{12}^2/(x_{13}^2 x_{23}^2)$ and the scaling properties of $t$ we can write
\beq
\langle \mathcal{O}_1(x_1 ,z_1)\mathcal{O}_2(x_2,z_2) \mathcal{O}_3(x_3 ,z_3)\rangle= \frac{t\left(\tilde{x}_{12},\tilde{z}_1,\tilde{z}_2 ,z_3\right)
}{x_{13}^{\Delta_1+\Delta_3-\Delta_2+\sum l_i}
x_{23}^{\Delta_2+\Delta_3-\Delta_1+\sum l_i}
x_{12}^{\Delta_1+\Delta_2-\Delta_3+\sum l_i}}\,,
\eeq
where
\beq
\tilde{x}_{12}= x_{13}^2 x_{23}^2 X_{12} =  x_{13} \,x_{23}^2 - x_{23}\,x_{13}^2\,.
\eeq

This expression treats the operator $\mathcal{O}_3$ differently from the other two operators. However, if needed, one can easily rewrite it,  so that the role 
of $\mathcal{O}_3$ is taken by, say, $\mathcal{O}_1$. To do this, one needs to re-express the numerator as
\beq
t\left(\tilde{x}_{12},R(x_{13}){z}_1,R(x_{23}){z}_2 ,z_3\right)=t'\left(\tilde{x}_{23},{z}_1,R(x_{12}){z}_2 ,R(x_{13})z_3\right)\,,
\eeq
where $t'$ is some other polynomial. To find $t'$, notice first of all that the transformation $R(x)$ is orthogonal.\footnote{It's actually a very trivial orthogonal transformation; it just flips the sign of the component in the direction of $x$.} Since $t$ is a rotationally invariant polynomial it will not change if every argument is multiplied by $R(x_{13})$. Using the relations
\beq
R(x_{13}) R(x_{23})=R(\tilde x_{23})R(x_{12}),\qquad R(x_{13}) \tilde x_{12}= \tilde x_{23}\,,
\eeq
we see that this transformation accomplishes the needed rewriting, and that
\beq
t'(x,z_1,z_2,z_3)=t(x,z_1,R(x)z_2,z_3)\,.
\eeq

Now, it is clear that if $\mathcal{O}_1=\mathcal{O}_2$, then the polynomial $t$ obeys
\beq
t(x,z_1,z_2,z_3)=t(-x,z_2,z_1,z_3)\,.
\eeq
On the other hand, if $\mathcal{O}_2=\mathcal{O}_3$ it is the $t'$ which satisfies the simple condition, while for $t$ the condition is less transparent:
\beq
t\left(x,z_1,z_2 ,z_3\right)=t\left(-x,z_1,R(x)z_3,R(x)z_2\right)\,.
\eeq

We now wish to compare with Eq.~(\ref{general3pt}).
In order to do that, we should project the embedding correlator onto the Poincar\'e section, using
\beq
P_i=(1,x_i^2,x_i)\ ,\ \ \ \ \ \ \ \ \ \ \ Z_i=(0,2x_i\cdot z_i,z_i)\ .
\eeq
One can then check that
\begin{align}
P_{23} V_1&=  -
 \tilde{z}_1\cdot \tilde{x}_{12} \ ,\ \ \ \ \ \ \ \ \ \ \ \ \ \ \ \ \ \ \,
P_{13} V_2=  - \tilde{z}_2\cdot \tilde{x}_{12} \ ,\ \ \ \ \ \ 
P_{12} V_3=   z_3 \cdot \tilde{x}_{12}\ , \nonumber\\
P_{12} P_{23}H_{13}& =( \tilde{z}_1\cdot z_3)\tilde{x}_{12}^2\ ,\ \ \ \ \ \ \ \ \ \ 
P_{12} P_{13}H_{23} = (\tilde{z}_2\cdot z_3)  \tilde{x}_{12}^2\ , \label{Emdtophysical}\\
P_{13} P_{23} H_{12}&=  (\tilde{z}_1\cdot \tilde{z}_2) \tilde{x}_{12}^2-2  (\tilde{z}_1\cdot \tilde{x}_{12})(\tilde{z}_2\cdot \tilde{x}_{12})\,. 
\nonumber
\end{align}
Therefore, the structure (\ref{Embterm}) corresponds to $t(x,z_1,z_2,z_3)$ given by
\beq
  (x^2 z_1\cdot z_3)^{n_{13}}(x^2 z_2\cdot z_3)^{n_{23}}
 (x^2 z_1\cdot z_2 -2x\cdot z_1\, x\cdot z_2 )^{n_{12}}\,
 (-x\cdot z_1)^{m_1} (-x\cdot z_2)^{m_2} (x\cdot z_3)^{m_3}\,,
\eeq
modulo terms $O(z_i^2)$ which are not independent but fixed by tracelessness of $\mathcal{O}$'s.
It is also clear that this is a basis for the most general  rotational and parity invariant polynomial $t(x,z_1,z_2,z_3)$.  

Parity odd structures are dimension specific. 
In order to form a scalar from the $d$-dimensional $\epsilon$-tensor we need at least $d$ linearly independent vectors.
Therefore, for $d>4$ the polynomial $t(x,z_1,z_2,z_3)$ is necessarily parity invariant, as stated in the previous section. In four dimensions, we can 
make parity odd three-point functions using $\epsilon( x, z_1,z_2,z_3)$.
This corresponds to the use of (\ref{epsilon(d=4)}) in the embedding language.
To see that, we just need to project onto the Poincar\'e section,
\beq
\epsilon(Z_1,Z_2,Z_3,P_1,P_2,P_3)=
\left|
\begin{array}{cccccc}
               0 & 0    & 0 &1 &1& 1     \\
               2z_1\cdot x_1    & 2z_2\cdot x_2  & 2z_3\cdot x_3 & x_1^2 & x_2^2 & x_3^2    \\
               z_1    & z_2 & z_3 & x_1 & x_2 & x_3
                         \end{array}\right|\,.
\eeq
Using translation invariance we can write  
\beq
\epsilon(Z_1,Z_2,Z_3,P_1,P_2,P_3)=
\left|
\begin{array}{cccccc}
               0 & 0    & 0 &1 &1& 1     \\
               2z_1\cdot x_{13}    & 2z_2\cdot x_{23}  &0 & x_{13}^2 & x_{23}^2 & 0   \\
               z_1    & z_2 & z_3 & x_{13} & x_{23} & 0
                         \end{array}\right|\,,
\eeq
and expanding in the last column, we find
\begin{align}
\epsilon(Z_1,Z_2,Z_3,P_1,P_2,P_3)
=\epsilon( \tilde{x}_{12},\tilde{z}_1,\tilde{z}_2,z_3)\ .
\end{align}

The problem in three dimensions is special so we treat it separately in the next subsection.

\subsubsection{Three Dimensions} \label{3D3pt}

The problem of constructing conformally invariant three-point functions in three dimensional CFTs has been recently addressed in \cite{Giombi}.  In this subsection we shall explain how their results fit into the formalism of this paper.
  
Using the group theoretic approach of \cite{Mack} it is easy to count how many independent structures exist for a three-point function of operators with spin $l_1\le l_2\le l_3$.
We just need to count how many irreducible representations of $SO(3)$ appear in the tensor product $l_1 \otimes l_2 \otimes l_3$ (notice that all irreducible representations of $SO(3)$ are totally symmetric and traceless representations).
This gives
\beq
N_{3d}(l_1,l_2,l_3)= \sum_{l=l_3-l_2}^{l_3+l_2}\sum_{m=|l-l_1|}^{ l+l_1}1
=(2l_1+1)(2l_2+1)-p(1+p)\,,
\eeq
where $p={\rm max} (0,l_1+l_2-l_3)$.
Of these, there are 
\begin{align}
N_{3d}^{+}(l_1,l_2,l_3)
 &= 2l_1 l_2 +l_1+l_2+1-\frac{p(p+1)}{2} \label{3dnumbereven}
 \end{align}
parity even structures and
\begin{align}
N_{3d}^{-}(l_1,l_2,l_3)&=  2l_1 l_2 +l_1+l_2-\frac{p(p+1)}{2}   
\label{3dnumberodd}
 \end{align}
parity odd structures.
The split between parity even and parity odd structures follows from the fact that in the product of two $SO(3)$ tensors with spin $l_1$ and $l_2$, the tensors with spin $l_1+l_2,l_1+l_2-2,\dots ,|l_1-l_2|$ are parity even, and the tensors with spin $l_1+l_2-1,l_1+l_2-3,\dots ,|l_1-l_2|+1$  are parity odd because they contain one $\epsilon$-tensor. 

The number of parity even structures $N_{3d}^+$ is smaller than the general result (\ref{N3pt}). To explain this mismatch we need to notice that, in three dimensions, there are identities relating some of the general tensor structures. The easiest way to derive these relations is to consider the expression for the leading OPE coefficient $t(x,z_1,z_2,z_3)$. 
As in section \ref{sec:tens}, we can restrict the polynomial to $z_i^2=0$, which translates to an $O(Z_i^2, Z_i\cdot P_i)$ term in the embedding space.

In three dimensions, the four arguments of $t$ cannot be linearly independent vectors:
\beq
x=\sum_{i=1}^3 \alpha_i \,z_i\,.
\eeq
For $z_i^2=0$, the coefficients $\alpha_i$ can be given explicitly as
\beq
\alpha_i= \frac{(z_j\cdot x)(z_k\cdot z_i)+(z_k\cdot x)(z_j\cdot z_i) - (z_i\cdot x)( z_j\cdot z_k )}{2 (z_i \cdot z_j )(  z_i \cdot z_k)}\ \ \ \ \ \ \ \ \ \ 
(j\neq k\neq i)\, .
\eeq
Another way to express the linear dependence is as
\beq
\left. \det_{1\le i,j \le 4}  (z_i\cdot z_j )\right|_{z_4=x} =0\,.
\label{detij=0}
\eeq
Using the rules in Eq.~(\ref{Emdtophysical}), this last identity corresponds to the relation
\beq
\left(V_1H_{23}+V_2H_{13}+V_3H_{12}+2V_1V_2V_3\right)^2\approx -2H_{12}H_{13}H_{23} 
\label{evenidentity}
\eeq
between the conformally invariant structures. Here $\approx$ means modulo $O(Z_i^2, Z_i\cdot P_i)$.
This identity is (the square of) the identity  (2.14) of  \cite{Giombi}.
The identity  (\ref{evenidentity}) can also be obtained directly from the $(3+2)$-dimensional embedding space by noting that the 6 vectors $Z_i$ and $P_i$ can not be linearly independent. Equation~(\ref{evenidentity}) then follows from
$\det_{1\le i,j \le 6}  (Z_i\cdot Z_j ) =0$ where $Z_{i+3} \to P_i$ for $i=1,2,3$.
The existence of this identity means that one does not need to use the substructure
$H_{12}H_{13}H_{23}$ to write the most general three-point function.
It is then simple to correct the overcounting of the general analysis for parity even structures, by subtracting all structures containing the factor $H_{12}H_{13}H_{23}$.
This gives
\beq
N_{3d}^{+}(l_1,l_2,l_3)=N (l_1,l_2,l_3)-N (l_1-2,l_2-2,l_3-2)\,,
\eeq 
which agrees with the counting (\ref{3dnumbereven}) from group theory.

We can also find relations between the parity odd structures by expanding the following determinant along the first line,
\begin{align}
\left|
\begin{array}{cccc}
               A_1 & A_2     &A_3 & \sum \alpha_iA_i     \\
               z_1    & z_2   &  z_3 & x
                         \end{array}\right|=0\ ,
\end{align}     
where we recall that $z_i$ and $x$ are three dimensional vectors here represented as columns. 
The simplest identity follows from choosing $A_i=x\cdot z_i$:
\beq
(x\cdot z_1)\, \epsilon(z_2,z_3,x) +(x\cdot z_2 )\, \epsilon(z_3,z_1,x)
+(x\cdot z_3)\, \epsilon(z_1,z_2,x)-x^2\, \epsilon(z_1,z_2,z_3)=0\ .
\eeq
This tells us that we never need to use the substructure $\epsilon(z_1,z_2,z_3)$, since it can be obtained as a linear combination of $\epsilon(z_i,z_j,x)$.  Furthermore, choosing
\beq
A_1=-(x\cdot z_1)^2\ ,\ \ \ \ \ \ \   
A_2=x^2\, (z_1\cdot z_2) -(x\cdot z_1)(x\cdot z_2)\,,\ \ \ \ \ \ \  
A_3=x^2\, (z_1\cdot z_3) -(x\cdot z_1)(x\cdot z_3)\,,
\eeq
we obtain
\beq
A_1\, \epsilon(z_2,z_3,x) -A_2\,\epsilon(z_1,z_3,x)+A_3\, \epsilon(z_1,z_2,x)=0\,,
\eeq
where we have used that $\sum \alpha_i A_i=0$ (as one can check from Eq.~(\ref{detij=0})).
This identity is invariant under the permutation $z_2 \leftrightarrow z_3$, but one can generate two more identities by permuting $z_1 \leftrightarrow z_2$ and
$z_1 \leftrightarrow z_3$.
In terms of our conformally invariant structures, these identities read
\begin{align}
0&\approx V_1^2\, \epsilon_{23}  +
(H_{12}+V_1V_2)\, \epsilon_{13}  
-(H_{13}+V_1V_3)\, \epsilon_{12} \,,  \nonumber \\
0&\approx V_2^2\, \epsilon_{13} 
+(H_{23}+V_2V_3)\, \epsilon_{12} + (H_{12}+V_1V_2)\, \epsilon_{23} \,, \nonumber  \\
0&\approx V_3^2\, \epsilon_{12}-(H_{13}+V_1V_3)\, \epsilon_{23}    +
(H_{23}+V_2V_3)\, \epsilon_{13}\,,  \label{oddidentities}
\end{align}
where
\beq
\epsilon_{ij} \equiv P_{ij}\,\epsilon(Z_i,Z_j, P_1,P_2,P_3)\ .
\eeq
This follows from the projections to the Poincar\'e section (\ref{Emdtophysical}) and 
\beq
\epsilon_{12}=- x_{12}^2 \,\epsilon(\tilde{z}_1,\tilde{z}_2,\tilde{x}_{12})    \ ,\ \ \ \ \ \ \ \ 
\epsilon_{13}=- x_{13}^2 \,\epsilon(\tilde{z}_1,z_3,\tilde{x}_{12})  \ ,\ \ \ \ \ \ \ \ 
\epsilon_{23}=- x_{23}^2 \,\epsilon(\tilde{z}_2,z_3,\tilde{x}_{12}) \ .
\eeq
The identities (\ref{oddidentities}) are equivalent to the Eqs.~(2.19) given in~\cite{Giombi}. The identity (\ref{evenidentity}) follows from the compatibility of these three equations.
The easiest way to count all parity odd three-point functions is to take these three identities as the only independent relations between the building blocks. 
Then we have
\begin{align}
N_{3d}^{-}(l_1,l_2,l_3)&=N (l_1-1,l_2-1,l_3)+N (l_1-1,l_2,l_3-1)+N (l_1 ,l_2-1,l_3-1) \nonumber\\
& \ \ \ -N (l_1-2,l_2-1,l_3-1)-N (l_1-1,l_2-2,l_3-1) \nonumber\\
& \ \ \ -N (l_1-1,l_2-1,l_3-2),
\end{align}
where the first line corresponds to all parity even structures times $\epsilon_{12}$, 
 $\epsilon_{13}$  and $\epsilon_{23}$, respectively.
 The second and third lines corresponds to the subtraction of the identities (\ref{oddidentities}), multiplied by parity even structures to avoid overcounting.
This expression agrees with the explicit formula given in Eq.~(\ref{3dnumberodd}).

\subsection{Four-Point Functions}
Now let us move on to discuss the possible structures that can appear in CFT four-point functions.  The simplest case is when all four operators are scalar primaries. A correlation function 
$\langle  \phi_1(x_1) \phi_2(x_2) \phi_3(x_3) \phi_4(x_4) \rangle$ containing primaries of dimension $\Delta_i$ can be obtained 
from the projection of the embedding correlator
\beq
\langle  \Phi_1(P_1) \Phi_2(P_2) \Phi_3(P_3) \Phi_4(P_4) \rangle =
 \left( \frac{P_{24}}{P_{14}}\right)^{\frac{\Delta_1 - \Delta_2}{2}} \left( \frac{P_{14}}{P_{13}}\right)^{\frac{\Delta_3 - \Delta_4}{2}} 
 \frac{ f(u,v)}{(P_{12})^{\frac{\Delta_1 + \Delta_2}{2}} (P_{34})^{\frac{\Delta_3 + \Delta_4}{2}}}\,,
\eeq
where $u$ and $v$ are the conformally invariant cross-ratios
\beq
u = \frac{P_{12} P_{34}}{P_{13} P_{24}} \,,\ \ \ \ \ \ \ \ \ \ \ 
v = \frac{P_{14} P_{23}}{P_{13} P_{24}} \,.
\eeq
Thus, in this very simple case, the correlation function depends on a single function of the cross ratios.

The generalization to operators with spin is clear and follows the same logic explained in section \ref{3pt}.
In this case, however,  the correlation function will be a linear combination of
tensor structures that are polynomial in the  $Z$'s, with coefficients
given by undetermined functions of the cross ratios. Thus, for a generic four-point function we write
\beq
\tilde G_{\chi_1,\chi_2,\chi_3,\chi_4} = 
 \frac{ \left( \frac{P_{24}}{P_{14}}\right)^{\frac{\tau_1 - \tau_2}{2}} \left( \frac{P_{14}}{P_{13}}\right)^{\frac{\tau_3 - \tau_4}{2}}}{(P_{12})^{\frac{\tau_1 + \tau_2}{2}} (P_{34})^{\frac{\tau_3 + \tau_4}{2}}}\,
  \sum_k  f_k(u,v) \,Q^{(k)}_{\chi_1,\chi_2,\chi_3,\chi_4}  (\{ P_i;Z_i\})\,,
  \label{4pf}
\eeq
where $\tau_i=\Delta_i+l_i$.
With this choice of pre-factor, the $Q^{(k)}$ have weight $l_i$ in each point $P_i$.
Conformal invariance is equivalent to the following condition for each linearly independent $Q^{(k)}$ polynomial:
\beq
Q^{(k)}_{\chi_1,\chi_2,\chi_3,\chi_4}(\{\lambda_i P_i; \alpha_i Z_i+\beta_i P_i\})=
 Q^{(k)}_{\chi_1,\chi_2,\chi_3,\chi_4}(\{P_i;Z_i\})\  
\prod_i \left(\lambda_i  \alpha_i \right)^{l_i} \,.
\label{Qconditions}
\eeq
Similar to the three-point function case, these polynomials are constructed from the basic building blocks $V_{i,jk}$ and $H_{ij}$
introduced in section \ref{3pt}. However, not all $V_{i,jk}$ are linearly independent. In addition to $V_{i,jk}=-V_{i,kj}$ we have, in the case of four points,
\beq
(P_2\cdot P_3)(P_1\cdot P_4) V_{1,23} +
 (P_2\cdot P_4)(P_1\cdot P_3) V_{1,42}+
 (P_3\cdot P_4)(P_1\cdot P_2) V_{1,34}=0\ .
\eeq
This shows that there are only 2 independent $V_{i,jk}$ for each $i$. 
A convenient choice for the example given below is to use linear combinations that are even and odd under the interchange $3 \leftrightarrow 4$,
 \begin{align}
W_1 &\equiv  V_{1,23} + V_{1,24}  \ , \ \ \ \ \ \ \ \ \ \  \
\bar{W}_1 \equiv V_{1,23} - V_{1,24}  \ , 
\label{V1s} \\ 
W_2 &\equiv  V_{2,13} + V_{2,14}  \ , \ \ \ \ \ \ \ \ \ \  \
\bar{W}_2 \equiv V_{2,13} - V_{2,14}  \ .
\end{align}
Similarly, we may define $W_3,W_4$ and $\bar{W}_3,\bar{W}_4$ to be, respectively, even and odd under 
the interchange $1 \leftrightarrow 2$.
Then, all solutions $Q^{(k)}$  of (\ref{Qconditions}) have the form
\beq
\prod_{i} W_i^{m_i} \prod_{i}\bar{W}_i^{\bar{m}_i} \prod_{i<j} H_{ij}^{n_{ij}}\,,
\eeq
such that
\beq
m_i + \bar{m}_i + \sum_{j\neq i} n_{ij}= l_i\,.
\eeq
The problem of finding the number of structures for the  four-point function is given by counting the 6-tuples 
$(n_{12},n_{13},n_{14},n_{23},n_{24},n_{34})$ of non-negative integers such that
\begin{align}
&n_{12}+n_{13}+n_{14} = a_1 \le l_1\,,\nonumber\\
&n_{12}+n_{23}+n_{24} = a_2 \le l_2\,,\nonumber\\ 
&n_{13}+n_{23}+n_{34} = a_3 \le l_3\,,\nonumber\\ 
&n_{14}+n_{24}+n_{34} = a_4 \le l_4\,.
\end{align}
Then, for each of these 6-tuples with a given set $\{a_i\}$, there are
\beq
\prod_{i=1}^4 (l_i-a_i+1)\,,
\eeq
possible ways of distributing the $W_i$ and $\bar{W}_i$ structures (counting number of integers $m_i$ and $\bar{m}_i$ such that $m_i+\bar{m}_i=l_i-a_i$).
We will not attempt here to count the number of general structures allowed for a generic  four-point function. The whole point of this analysis was to make it clear how to construct such structures in any given particular case that one may wish to consider.

\subsubsection{Example: Vector-Vector-Scalar-Scalar}

As an example of the previous general formalism let us consider the case of a four-point function between two vectors
and two scalars
$\langle v_a(x_1) v_b(x_2) \phi(x_3) \phi(x_4) \rangle$, even under the exchange of both vectors and of both scalars.
To start there are five possible independent structures, namely
\beq
W_1W_2\,,\ \  \bar{W}_1 \bar{W}_2\,,\ \  W_1 \bar{W}_2,\ \   \bar{W}_1W_2, \ \  H_{12}\,.
\eeq
Noticing that under $P_1\leftrightarrow P_2$ or $P_3\leftrightarrow P_4$ the cross ratios transform as $u \leftrightarrow  w\equiv u/v$,
it is clear that in this case the linear combination of the $Q^{(k)}$ entering (\ref{4pf}) is given by
\beq
f_1(u,w) W_1W_2 + f_2(u,w) \bar{W}_1 \bar{W}_2 +  f_3(u,w)H_{12} +  f_4(u,w) \left( W_1 \bar{W}_2 - \bar{W}_1W_2 \right)\,,
\eeq
with
\beq
f_4(u,w) = -f_4(w,u)\ ,\ \ \ \ \ \ \ \ f_k(u,w) = f_k(w,u)\,,\ \ \ k=1,2,3\ .
\eeq
Hence we recover the counting already presented in \cite{CCP}.

\subsection{$n$-Point Functions}

We will finish this section with some general remarks on the case of $n$-point functions, for which
there are $n(n-3)/2$  independent conformally invariant cross-ratios $u_a$ (actually, for $n$ high enough
they are  not all independent, but this fact will not be important here).

A generic $n$-point function can be written as
\beq
\tilde G_{\chi_1,\dots,\chi_n} =  \prod_{i<j}^n P_{ij}^{-\alpha_{ij}} \,
\sum_k  f_k(u_a) \,Q^{(k)}_{\chi_1, \dots ,\chi_n}  (\{ P_i;Z_i\})\,,
\label{npf}
\eeq
where 
\beq
\alpha_{ij}=\frac{ \tau_i+\tau_j}{n-2} - \frac{1}{(n-1)(n-2)} \sum_{k=1}^n \tau_k\ .
\label{alpha_ij}
\eeq 
With the chosen pre-factor, the $Q^{(k)}$ have weight $l_i$ in each point $P_i$. They are also identically transverse:
\beq
Q^{(k)}_{\chi_1,\dots,\chi_n}(\{\lambda_i P_i; \alpha_i Z_i+\beta_i P_i\})=
 Q^{(k)}_{\chi_1,\dots,\chi_n}(\{P_i;Z_i\})\  
\prod_i \left(\lambda_i  \alpha_i \right)^{l_i} \,.
\eeq
These polynomials can then be constructed from the basic building blocks $V_{i,  jk} $ and $H_{ij}$ given in  (\ref{V_i,jk}) and (\ref{H_ij}).
For each $i$, since only $n-2$ of the (anti-symmetric) $V_{i,jk}$ are linearly independent, we can choose
to work with
\beq
\mathcal{V}_{ij} \equiv V_{i, (i+1) j}\ \ \ \ \ \ \ \ \ \ \ \ \ \ (j=1,\cdots,\hat{i},\hat{i+1},\cdots, n)\,,
\eeq
where hatted integers are excluded.
Then, all solutions $Q^{(k)}$   have the form
\beq
\left(\prod_{i=1}^n  \prod_{j\neq i,i+1}^n \mathcal{V}_{ij}^{m_{ij}} \right)
\prod_{i<j}^n H_{ij}^{n_{ij}}\,,
\eeq
such that
\beq
\sum_{j\neq i,i+1}^n  m_{ij} + \sum_{j\neq i}^n  n_{ij}= l_i\,. \label{nptcounting}
\eeq
Thus, the problem of finding the number of structures of the $n$-point function 
separates again in finding the $(n(n-1)/2)$-tuples, $\{ n_{ij}\}$ with $i<j$,  such that
\beq
\sum_{j\neq i}^n  n_{ij}=a_i\le  l_i \label{nijai}\,.
\eeq
For each set of non-negative integers $a_i$, a moment's  thought shows that there are
\beq
\prod_{i=1}^n \frac{ (l_i-a_i +n -3)!}{ (l_i-a_i)! (n-3)!}
\eeq
possible ways of distributing the $\mathcal{V}_{ij}$ structures.

In the above counting we neglected identities following from the finite dimensionality of spacetime. The $2n$ vectors $Z_i$ and $P_i$ can not be linearly independent in the $(d+2)$-dimensional embedding space if $n>\frac{d}{2}+1$. In a given dimension, one can obtain identities between the above tensor structures by expanding 
${\rm det} (Z_i\cdot Z_j)=0$, where the matrix is of size $(d+3)\times(d+3)$ or larger and some of the $Z$'s can be $P$'s.
  
  \begin{figure}[htbp]
\begin{center}
\includegraphics[scale=0.4]{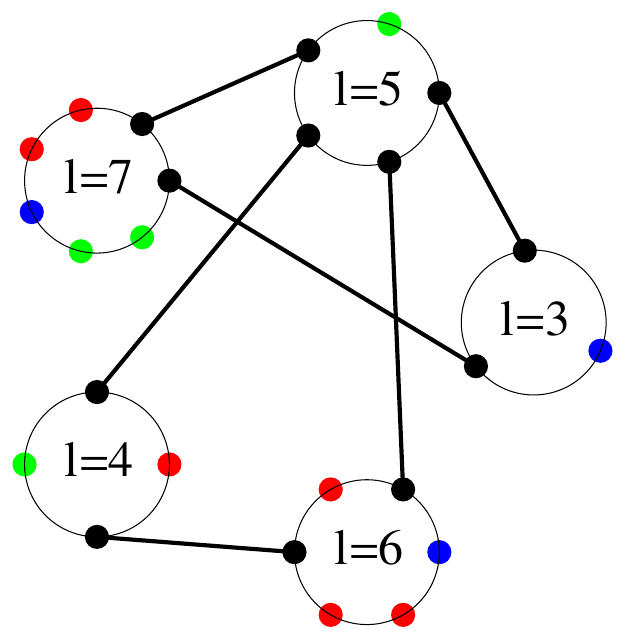}
\caption{Same as Fig.~\ref{Diagram3pt} but for a five-point function. The isolated dots representing $V$'s appear in several colors because for an $n$-point function there are several possible $V$'s per vertex.}
\label{Diagram5pt}
\end{center}
\end{figure}

\section{Conserved Tensors}
\label{sec:cons} 

In unitary CFTs, the dimensions of spin $l$ primaries must satisfy the unitarity bound \cite{unitarity4D,unitarity}:
\beq
\Delta\ge l+d-2\quad(l\ge1)\,.\label{UB}
\eeq
When $\Delta$ takes the lowest value allowed by this bound for a given $l$, the corresponding primary field is conserved. Physically important examples of such fields are the stress tensor ($l=2$) and global symmetry currents ($l=1$).\footnote{Note that it is not as interesting to consider scalars, since only a free field can saturate the scalar unitarity bound $\Delta\ge (d-2)/2$.} The conservation condition then leads to additional constraints on the form of three and higher point functions. In this section we will discuss these constraints and show how to impose them directly in the embedding space.

\subsection{Conservation Condition and Conformal Invariance} 
\label{CC}
Let us begin by considering the conservation condition for a spin $l$ dimension $\Delta$ primary:
\begin{gather}
\partial \cdot f=0\, ,
\label{conserved}\\
(\partial \cdot f)^{a_2\ldots a_l}\equiv \frac{\partial}{\partial x^{a_1}} f^{a_1 a_2\ldots a_l}(x) \,.
\end{gather}
We would like to learn how to impose this condition in terms of the embedding space tensor $F$ which projects to $f$. Differentiating Eq.\ (\ref{proj}), there will be two types of terms depending whether the derivative falls on $\partial P/\partial x$ or on $F$. These terms can be simplified using
\begin{gather}
{\del \over \del x^a}\left({\del P^A \over \del x^b}\right)=\delta_{ab}\bar P^A , \\
{\del P^{A_1} \over \del x_{a_1}}{\del F_{A_1\ldots A_l} \over \del x^{a_1}}=
{\del P^{A_1} \over \del x_{a_1}}{\del P^{B} \over \del x^{a_1}}{\del F_{A_1\ldots A_l} \over \del P^B}\equiv K^{A_1B}{\del F_{A_1\ldots A_l} \over \del P^B}\,,
\end{gather}
where the metric $K^{AB}$ and the vector $\bar P^A$ were given in Eq.\ (\ref{id1}). Commuting $P$ with $\del/\del P$ and using the property that $F$ is homogeneous of degree $-\Delta$, the end result can be put in the form
\begin{equation}
(\partial \cdot f)_{a_2\ldots a_l}(x) = {\del P^{A_2} \over \del x^{a_2}} \ldots {\del P^{A_l} \over \del x^{a_l}} R_{A_2\ldots A_l}(P_x) \,,
\end{equation}
with
\begin{equation}
R_{A_2\ldots A_l}(P)=
\left[
\frac{\del}{\del P_{A_1}}-\frac{1}{P\cdot \bar P}
(\bar P\cdot \frac\del{\del P}) P^{A_1}-(l+d-2-\Delta)\frac{\bar P^{A_1}}{P\cdot \bar P}
\right] F_{A_1\ldots A_l}(P)\,.
\label{divproj}
\end{equation}
Note that the ${1}/{(P\cdot \bar P)}$ prefactors are needed to ensure that all terms in $R$ have the same homogeneity in $P$.

The tensor $F$ is originally defined on the cone $P^2=0$, while the derivatives $\del/\del P$ appearing in the definition of $R$ are unrestricted. To compute the derivatives along the non-tangent directions, the tensor $F$ has to be extended away from the cone. It is easy to see that different extensions of $F$ change $R$ by terms which project to zero. This is a sanity check, since the LHS of the formula does not allow for any ambiguity. The same is true about pure gauge modifications of $F$. 

The terms in $R$ involving $\bar P$ may seem problematic from the point of view of \CG\ invariance. The last term clearly breaks it unless its coefficient vanishes. On the other hand, the second term \emph{is} \CG\ invariant, though not manifestly. To see this, one should use the condition that $P\cdot F$ vanishes on the cone. Writing this as $P\cdot F=O(P^2)$, we see that $P\cdot \bar P$ cancels out and $\bar P$ drops out from the second term.

Now we see what is special about $\Delta=l+d-2$: precisely for this dimension $R$ becomes an \CG\ invariant tensor. This tensor is also traceless (obvious) and transverse (straightforward to show by using the tracelessness and transversality of $F$). We conclude that its projection to the physical space, $\del \cdot f$, will transform as a primary under the conformal group. In particular, the transformation of $\del \cdot f$ will be homogeneous: $\del \cdot f(x)$ is proportional to $\del \cdot f(x')$. This is to be contrasted with the usual transformation rule for the derivative of a primary, which contains a term proportional to the primary itself. 

One consequence of the above discussion is that for $\Delta=l+d-2$, and only for this dimension, the conservation condition $\del\cdot f=0$ can be imposed in a way that is consistent with the conformal symmetry.
 
 But one can say more. The fact that for $\Delta=l+d-2$ the divergence $\del \cdot f$ is both a primary and a descendant implies, using the argument familiar from 2D CFT, that it is a \emph{null state}. In particular, the two-point function of $\partial f$ with itself, as with any other primary, will vanish:
 \beq
 \langle \del\cdot f(x)\  \del\cdot f(0)\rangle=0\,.
 \label{null}
 \eeq
 The latter equality can be also checked using the two-point function of spin $l$ primaries discussed in section \ref{2pt}.
 
 Now, in a unitary theory Eq.\ (\ref{null}) implies that $\del\cdot f=0$ as an operator equation. Thus imposing the conservation condition for $\Delta=l+d-2$ is not only consistent, but also mandatory.
 
 In practice, we will have to impose that three-point functions of $f$ with any other fields should be conserved. However, unlike for the two-point functions, this will not happen automatically. Rather, we will find constraints beyond those discussed in section \ref{3pt}. On the other hand, once all the three-point function constraints are satisfied, higher point functions will be automatically conserved as a consequence of the OPE.
 
\subsection{Conservation Condition for Polynomials} 

Since the conservation constraint must be imposed in addition to the constraints discussed in section \ref{sec:corr}, we should write it in a form compatible with the index-free notation that we developed there. In particular, we will work with the encoding polynomial $\tilde F(P,Z)$ introduced in section \ref{sec:tens.emb.}, which is identically transverse and agrees with $F(P,Z)$ modulo $O(Z^2,Z\cdot P)$. Similarly, we will also encode the tensor $R$ via the identically transverse function $\tilde R(P,Z)$.

The result of this section will be that $\tilde R(P,Z)$ can be computed from $\tilde F(P,Z)$ by the following simple formula:
\beq
\tilde R(P,Z)=\frac{1}{l(h+l-2)}(\partial\cdot D) \tilde{F}(P,Z) -O(Z^2,Z\cdot P) ,
\label{miracle}
\eeq
where 
\beq
\partial\cdot D\equiv \frac{\partial \ }{\partial P_M} \,D_M ,
\eeq
and $D_M$ is the differential operator in $Z$ defined in Eq.\ (\ref{EmbTod}). ``$-O(\cdots)$" means that the corresponding terms must be dropped.

\vspace{1em}
{\proof Let us prove formula (\ref{miracle}). First we need to recover $F$ from $\tilde F$. According to the result from section \ref{sec:tens.emb.}, the necessary projector can be obtained from a $d$-dimensional traceless symmetric projector:
\beq
\pi_{a_1\ldots a_l,b_1\ldots b_l}=\delta_{a_1b_1}\cdots \delta_{a_lb_l}
-c_l \sum_{i<j}\delta_{a_ia_j}\delta_ {b_i b_j}\prod_{k\ne i,j}\delta_{a_kb_k}
+O(\delta_{a_ia_j}\delta_{a_ka_n}) \,.
\eeq
Here we are not symmetrizing in $b$'s, assuming that $\pi$ is contracted with a symmetric tensor. The second term in the formula subtracts single traces, which fixes its coefficient $c_l=1/(d+2l-4)$. The $O(\delta_{a_ia_j}\delta_{a_ka_n})$ stands for terms which subtract multiple traces; we will not need to know them explicitly. Performing the replacements from Eq.~\reef{repl1}, we obtain the representation
\beq
F_{A_1\ldots A_l}=\tilde F_{A_1\ldots A_l}
-c_l \sum_{i<j}W_{A_iA_j} \tilde F^B_{\ BA_1\ldots \hat A_i\ldots \hat A_j \ldots A_l}+O(W_{A_iA_j}W_{A_kA_n}) \,,
\eeq
where the hatted indices are skipped.

Now we can start computing $\tilde R$. Assuming that $\Delta=l+d-2$, Eq. (\ref{divproj}) gives
\beq
\tilde R_{A_2\ldots A_l}=
\left[
\frac{\del}{\del P_{A_1}}-\frac{1}{P\cdot \bar P}
(\bar P\cdot \frac\del{\del P}) P^{A_1}
\right] F_{A_1\ldots A_l}-O(\eta_{A_i A_j},P_{A_i})\, ,
\eeq
where $-O(\cdots)$ again indicates the terms which will be dropped when passing from $R$ to $\tilde R$. In fact, it is easy to see that the $O(W_{A_iA_j}W_{A_kA_n})$ part of $F$ only leads to such terms.  Similarly, all of the terms in $F$ proportional to $\eta_{A_i A_j}$ with $i,j \ne 1$ will also be dropped.

The remaining terms are
\beq
F_{A_1\ldots A_l} = \Bigl(\tilde F_{A_1\ldots A_l}
-c_l \sum_{j\ge 2}\eta_{A_1A_j} \tilde F^B_{\ BA_2\ldots \hat A_j \ldots A_l} \Bigr)+ c_l \sum_i P_{A_i} S_{A_1\ldots \hat A_i\ldots A_l} + \cdots\,,
\label{eq:Frem}
\eeq
where
\beq
S_{A_2 \ldots A_l}\equiv \frac1{P\cdot \bar P} \sum_{j=2}^l \bar P_{A_j}\tilde F^B_{\ BA_2\ldots \hat A_j\ldots A_l} \,.
\label{eq:S}
\eeq

Now let us apply the differential operator.  Using the fact that $\tilde F$ is transverse, the action on the first term of Eq. (\ref{eq:Frem}) gives
 \beq
\tilde R_{A_2\ldots A_l}= \frac{\del}{\del P_{A_1}} \Bigl( \tilde F_{A_1\ldots A_l}
-c_l \sum_{j\ge 2}\eta_{A_1A_j} \tilde F^B_{\ BA_2\ldots \hat A_j \ldots A_l} \Bigr) + c_l S_{A_2 \ldots A_l} - O(\eta_{A_i A_j},P_{A_i}) + \cdots\,,
\label{eq:contr1}
\eeq
where the $-O(\cdots)$ reminds us that some of the terms generated by ${\del}/{\del P_{A_1}}$ will have to be dropped.

To compute the action on the second term, we use a formula valid for any $S$ of homogeneity $-\Delta_S$:
\begin{multline}
\left[
\frac{\del}{\del P_{A_1}}-\frac{1}{P\cdot \bar P}
(\bar P\cdot \frac\del{\del P}) P^{A_1}
\right]\sum_i P_{A_i} S_{A_1\ldots \hat A_i\ldots A_l}= \\
=(d+ l-1-\Delta_S) S_{A_2\ldots A_l} -\frac{1}{P\cdot \bar P}\sum_i \bar P_{A_i} (P\cdot S)_{A_2\ldots \hat A_i\ldots A_l}+O(\eta_{A_i A_j},P_{A_i}) \,.
\end{multline}
Specializing to the $S$ in Eq. \reef{eq:S}, $\Delta_S=d+l-1$ and the first term vanishes. Using the contraction
\beq
(P\cdot S)_{A_3\ldots A_l}=\tilde F^B_{\ BA_3\ldots  A_l}
\eeq
(for $\tilde F$ transverse), we see that the contribution to $\tilde R_{A_2 \ldots A_l}$ is simply $- c_l S_{A_2 \ldots A_l}$, canceling the second term in 
Eq. (\ref{eq:contr1}).  Thus we obtain the final result
\beq
\tilde R_{A_2\ldots A_l} = \frac{\del}{\del P_{A_1}}  \Bigl( \tilde F_{A_1\ldots A_l}
-c_l \sum_{j\ge 2}\eta_{A_1A_j} \tilde F^B_{\ BA_2\ldots \hat A_j \ldots A_l} \Bigr) -O(\eta_{A_i A_j},P_{A_i})\, .
\eeq
It remains to convert this equation to the polynomial notation by contracting with $Z$'s. Using the definition of the operator $D_M$, it is straightforward to show that the resulting formula is identical to Eq.~\reef{miracle}.}

\subsection{Examples} 
\label{sec:CE}

Now we will give some simple examples of how to apply the above formalism, focusing on three-point functions where two of the three operators are conserved currents. We will then show how the conservation condition restricts possible structures that appear in these three-point functions.  Conservation constraints on the structure of three-point functions have been studied previously by Osborn and Petkou \cite{OP}, directly in the physical space. Where comparison is possible, we have verified explicitly that our methods reproduce their results. We consider only the parity even case in $d\ge 4$.

Let us consider the simplest nontrivial example of a three-point function between two vector currents at points $x_1$ and $x_2$ and a scalar operator at $x_3$, 
\beq
\langle  v^1_a(x_1) v^2_b(x_2)  \phi (x_3) \rangle\, .
\eeq
Here we assume that $\phi$ has dimension $\Delta$, while $v$'s necessarily have dimension $d-1$. The currents do not necessarily belong to the same nonabelian current multiplet, so we can consider both symmetry possibilities under the exchange of $v$'s. 

First we consider the symmetric case (e.g.~if the currents are identical). According to the results of section \ref{3pt}, the embedding function encoding this three-point function has the form
\beq
\tilde G(P_1, P_2, P_3; Z_1, Z_2) = \frac{  \alpha V_1V_2 + \beta H_{12} }{\left(  P_{12} \right)^{d - \frac{\Delta}{2} }\left(  P_{13}\right)^{\frac{\Delta}{2}  }\left( P_{23} \right)^{ \frac{\Delta}{2} }}\,,
\label{VVS}
\eeq
with a priori independent constants $\alpha$ and $\beta$. The conservation condition can be imposed by using Eq.~\reef{miracle}. Computing the divergence at $P_1$ and dropping the terms of $O(Z_1^2, Z_1\cdot P_1)$, we find the result
\beq
(\partial_{P_1}\cdot D_{Z_1})\,\tilde G \to
\left( \frac{d}{2} - 1 \right) (\alpha  (d- 1 - \Delta ) + \beta \Delta) 
 \frac{  V_2}{\left(  P_{12} \right)^{d - \frac{\Delta}{2} }\left(  P_{13}\right)^{\frac{\Delta}{2}  }\left( P_{23} \right)^{ \frac{\Delta}{2} }}\,.
\eeq
For any $\alpha$ and $\beta$, this embedding function is identically transverse and has the correct structure to represent a three-point function between a scalar $\partial^a v^1_a(x_1)$, a vector $v^2_b(x_2)$ and another scalar  $\phi(x_3)$. This is exactly how it should be, since taking divergence is consistent with conformal symmetry for the canonical field dimensions. Moreover, current conservation demands that the result should actually vanish, which implies that $\alpha$ and $\beta$ must be related by
\beq
 \alpha  ( d-1-\Delta) + \beta \Delta =0 .
\eeq

This example demonstrates how the conservation condition can be simply imposed directly in the embedding space. Note that the computations in this formalism are completely mechanical and easily lend themselves to automatization, e.g. in {\sc Mathematica}. 
 
Let us now generalize to the three-point function when the scalar is replaced by a spin $l$, dimension $\Delta$ operator:
\beq
\langle  v^1_a(x_1) v^2_b(x_2)  {\cal O}_{c_1\cdots c_l}(x_3) \rangle,
\label{vvl}
\eeq
still symmetric in $1\leftrightarrow 2$. When $l\ge 2$ is even this three-point function has an embedding function that a priori depends on the four constants $\alpha$, $\beta$, $\gamma$ and $\eta$:
\beq
\tilde G(\{P_i; Z_i\}) = 
\frac{ \alpha V_1V_2 V_3^l   + \beta \left( H_{13}  V_2 + H_{23} V_1  \right) V_3^{l-1} + \gamma H_{12} V_3^l +  \eta H_{13} H_{23} V_3^{l-2} }{\left(  P_{12} \right)^{d-\frac{\Delta+l}{2} }\left(  P_{13}\right)^{\frac{\Delta+l}{2} }\left( P_{23} \right)^{\frac{\Delta+l}{2} }} .
\label{VVL}
\eeq
The particular combinations of elementary tensor structures are fixed by the requirement that the function be even under the exchange of points $P_1$ and $P_2$. Computing the divergence at $P_1$ and dropping the usual terms, we find the following result
\beq
(\partial_{P_1}\cdot D_{Z_1})\,\tilde G \to  \left( \frac{d}{2} - 1 \right) 
\frac{ a V_2 V_3^l   + b H_{23} V_3^{l-1}  }{\left(  P_{12} \right)^{d-\frac{\Delta+l}{2} }\left(  P_{13}\right)^{\frac{\Delta+l}{2} }\left( P_{23} \right)^{\frac{\Delta+l}{2} }}\,,
\eeq
with
\begin{align}
a& = \alpha (d - 1- \Delta ) + \beta (2 - 2 d - l + \Delta) +  \gamma (l + \Delta)\,,
\\
b&=  \beta (  d-2  - \Delta)  + \gamma l +  \eta (4 - 2 d - l + \Delta)\,.
\end{align}
Current conservation then forces $a=b=0$, reducing the number of independent tensor structures in this three-point function from four to two.

For odd $l\ge 1$ there is a single tensor structure invariant under the exchange of points $P_1$ and $P_2$, given by
\beq
V(\{P_i; Z_i\}) = 
\frac{ \alpha  \left( H_{13}  V_2 - H_{23} V_1  \right) V_3^{l-1}  }{\left(  P_{12} \right)^{d-\frac{\Delta+l}{2} }\left(  P_{13}\right)^{\frac{\Delta+l}{2} }\left( P_{23} \right)^{\frac{\Delta+l}{2} }}\,.
\eeq
However, imposing conservation as above, we find $\alpha=0$. This means that an odd $l$ field cannot appear in the OPE of two identical conserved currents.

The three-point functions anti-symmetric under current exchanges are straightforward to consider by the same method.
One can also consider (spin 2)-(spin 2)-(spin $l$) three-point function, imposing stress tensor conservation (appendix~\ref{app:a}). The results are summarized in table 1.
\begin{table}
\centering
\begin{tabular}{| l | c | c |}
\hline
 						& symmetric  & anti-symmetric \\ \cline{2-3}
$\langle v^1 v^2 O^{(l)}\rangle$  & $\begin{array}{rcl} l=0&:& 2\to \boldsymbol{1} \\ 
                                                  l\ge 1\text{ odd }&:& 1\to\boldsymbol{0} \\
                                                  l\ge 2\text{ even }&:&4\to\boldsymbol{2} 
                                                 \end{array}$ & 
                                                   $\begin{array}{rcl}  l=0 &:& 0 \\ 
                                                  l= 1\text{ conserved }&:& 3\to\boldsymbol{2}\ \text{\cite{Schreier}}\\
                                                  l=1\text{ non-conserved }&:&3\to\boldsymbol{1} \\
                                                   l\ge2\text{ even }&:& 1\to\boldsymbol{0} \\ 
                                                  l\ge 3\text{ odd }&:&4\to\boldsymbol{2} 
                                                                                                   \end{array}$ \\
                                 \hline    \hline
                                  					
$\langle T T O^{(l)}\rangle$  & \multicolumn{2}{|c|}{
						$\begin{array}{rcl}  l=0\  &:&3\to \boldsymbol{1}\ \text{\cite{OP}}\\ 
                                                  l\ge 1\text{ odd }&:&4\to\boldsymbol{0} \\
                                                   l= 2\text{ conserved }&:&8\to\boldsymbol{3} \\
                                                   l= 2\text{ non-conserved }&:& 8\to\boldsymbol{2} \\
                                                  l\ge 4\text{ even }&:&10\to\boldsymbol{3} 
                                                 \end{array}$} \\
                                                 \hline
\end{tabular}
\caption{The number of parity even structures in the three-point function of two conserved spin $j$ currents ($j=1,2$) with an arbitrary spin $l$ primary in $d\ge 4$. We consider symmetric and anti-symmetric structures with respect to exchanging spin 1 currents, while only symmetric structures are relevant for the stress tensor correlators. ``$n\to \boldsymbol{m}$" means that $n$ conformal structures compatible with the assumed exchange symmetry are reduced to $\boldsymbol{m}$ when the conservation condition is imposed.}
\end{table}

We would like to comment about the case when the spin $l$ operator is also conserved. One could na\"ively expect that imposing spin $l$ conservation would lead to a further reduction of structures, but that's not what happens. For $l$ unequal to the spin $j$ of the other two conserved currents, spin $l$ conservation turns out to be satisfied automatically as a consequence of the spin $j$ conservation and setting the spin $l$ dimension to the canonical value $\Delta=l+d-2$. Furthermore, for $l=j$ we actually get one more structure by going to the canonical spin $l$ dimension, as the table shows. What happens is that for this dimension some of the constraints for the coefficients of elementary structures become linearly dependent.

\section{S-Matrix Rule for Counting Structures}
\label{sec:S} 
 
 In the previous sections we have rigorously derived a number of results related to counting CFT three-point function structures, with or without conservation constraints. We will now present a rule which allows us to intuitively explain all of the found results. 
The first appearance of this rule was the observation by Hofman and Maldacena \cite{Hofman} that the number of conformally invariant structures in the stress tensor three-point function in $d\ge4$, computed to be 3 by Osborn and Petkou \cite{OP}, coincides with the number of on-shell three-graviton vertices in $\bM^{d+1}$, computed to be 3 by Metsaev and Tseytlin \cite{Metsaev}. 

We propose the following generalization of this rule, which covers both the conserved and non-conserved case: \emph{The number of independent structures in a three-point function containing operators of spins $\{l_1,l_2,l_3\}$ is equal to the number of independent on-shell scattering amplitudes for particles of spins $\{l_1,l_2,l_3\}$ in $d+1$ flat Minkowski dimensions. The particles should be taken massless or massive depending on whether or not the corresponding operator is conserved.}

To demonstrate how this works, let us first consider the case of a scattering amplitude between 3 massive particles of arbitrary spin. It is a Lorentz invariant function of the momentum $p_i$ and polarization tensor $\zeta_i$ of each particle.
Since the spin $l_i$ polarization tensors $\zeta_i$ are symmetric and traceless, we can trade them for a polynomial of degree $l_i$ in the null vector $z_i$.
Moreover, the transversality condition 
$  (p_{i})_{ \mu_1}\zeta_i^{\mu_1\dots\mu_{l_1}} =0$ translates to $z_i\cdot p_i=0$.\footnote{That this is the right condition to recover the tensor is clear in the rest frame of the particle, where the polarization tensor is purely spatial.}
Therefore, we must count polynomials such that
\beq
S(p_1,p_2,p_3;\lambda_1 z_1,\lambda_2 z_2,\lambda_3z_3)=
\lambda_1^{l_1}\lambda_2^{l_2}\lambda_3^{l_3}
S(p_1,p_2,p_3; z_1,  z_2, z_3)\,,
\eeq 
where $z_i\cdot p_i=0$ and 
\beq
p_1+p_2+p_3=0\ ,\ \ \ \ \ \ \ \ p_i^2=-M_i^2\ .
\eeq
On-shellness and momentum conservation tell us that the contractions $p_i\cdot p_j$ can be written in terms of the particle masses and can therefore be dropped.
Further, momentum conservation and transversality imply that 
$z_1\cdot p_2= -z_1\cdot p_3$. Therefore, the general solution is a linear combination of 
\beq
S(n_{12},n_{13},n_{23})=(z_1\cdot z_2)^{n_{12}}(z_1\cdot z_3)^{n_{13}}
(z_2\cdot z_3)^{n_{23}}
(z_1\cdot p_2)^{m_1}
(z_2\cdot p_3)^{m_2}
(z_3\cdot p_1)^{m_3} \,,
\label{3ptSmatrices}
\eeq
where
\beq
m_i  =l_i-\sum_{j\neq i} n_{ij}\ge 0\,. 
\eeq
Since this is the same condition as Eq.~(\ref{expconstraints}), the number of solutions is given by exactly the same combinatorial problem that we solved for CFT three-point functions.
It is clear that there are no parity odd structures available in dimension bigger than 5 and in 5 dimensions we have the unique structure
\beq
\epsilon(z_1,z_2,z_3,p_1,p_2)=-\epsilon(z_1,z_2,z_3,p_1,p_3)
=\epsilon(z_1,z_2,z_3,p_2,p_3)\ , 
\eeq
in perfect agreement with the results of section \ref{parityodd} for parity odd correlators.

Actually,  the rule seems to work even beyond the three-point function level. Indeed, the most general $n$-particle scattering amplitude is a linear combination of 
\beq
 \left(\prod_{i=1}^n  \prod_{j\neq i,i+1}^n (z_i\cdot p_j)^{m_{ij}} \right)
\prod_{i<j}^n (z_i\cdot z_j) ^{n_{ij}}\,,
\eeq
where
\beq
\sum_{j\neq i,i+1}^n  m_{ij} + \sum_{j\neq i}^n  n_{ij}= l_i\,.
\eeq
This is identical to the condition (\ref{nptcounting}) for counting general tensor structures in an $n$-point conformal correlator.
Moreover, the coefficients in the linear combination of structures for the S-matrix 
can be arbitrary functions
of the Mandelstam invariants, in direct analogy with 
 the functions $f_k $ of the cross-ratios in the $n$-point conformal correlators (\ref{npf}).
This match strongly suggests that there is a one-to-one correspondence between 
 $n$-particle scattering amplitudes and $n$-point conformal correlators.

\subsection{Massless Particles}
\label{massless} 
  
Let us now study massless particles.
In this case, the scattering amplitude must be invariant under the infinitesimal gauge transformation
\beq
\zeta_{\mu_1 \dots \mu_l} \to 
\zeta_{\mu_1 \dots \mu_l} + p_{(\mu_1}\Lambda_{\mu_2\dots\mu_l)}\ .
\eeq
This corresponds to invariance under
\beq
z_\mu \to z_\mu+  \epsilon \,p_\mu \label{gaugetransf}
\eeq
to first order in $\epsilon$.
The problem of finding gauge invariant 3-particle scattering amplitudes is then reduced to finding linear combinations of the structures (\ref{3ptSmatrices}) that are invariant under \reef{gaugetransf} to first order in $\epsilon$.
Recalling that $p_i^2=0$, it is easy to see that
\begin{align}
\delta_1 S(n_{12},n_{13},n_{23}) &=
\epsilon_1 \Big[  
n_{13}\, S_1(n_{12},n_{13}-1,n_{23}) -n_{12}  \,S_1(n_{12}-1,n_{13},n_{23}) \Big]\,,
\label{gaugecondition1}\\
\delta_2 S(n_{12},n_{13},n_{23}) &=
\epsilon_2 \Big[  
n_{12} \,S_2(n_{12}-1,n_{13},n_{23}) -n_{23}  \,S_2(n_{12},n_{13},n_{23}-1) \Big]\,,
\label{gaugecondition2}\\
\delta_3 S(n_{12},n_{13},n_{23}) &=
\epsilon_3 \Big[    n_{23} \, S_3(n_{12},n_{13},n_{23}-1)
-n_{13}\, S_3(n_{12},n_{13}-1,n_{23})  \Big]\,,
\label{gaugecondition3}
\end{align}
where $S_i$ is given by the same expression as $S$ but with 
$l_i \to l_i-1$.
This suggests starting with the ansatz
\beq
\sum_{i=0}^k a_i \, S(i,k-i,n_{23})
\eeq
to impose gauge invariance for particle 1.
We then find that
\begin{align}
0&=\sum_{i=0}^k \left(a_i \,i \, S_1(i-1,k-i,n_{23})-a_i \,(k-i) \, S_1(i,k-i-1,n_{23})\right) \nonumber\\
&=\sum_{i=1}^k \left(a_i \,i-a_{i-1} \,(k-i+1) \right) \, S_1(i-1,k-i,n_{23})\,,
\end{align}
which fixes all the coefficients up to an overall normalization,
\beq
a_i= \frac{k-i+1}{i} a_{i-1}=\frac{k!}{i!(k-i)!} a_0 \ .
\eeq
Notice that this solution only exists for $k \le l_1$.

Imposing gauge invariance also on particle 2, we find the amplitude
\beq
T_k=\sum_{i=0}^k \sum_{j=0}^{k-i} \frac{k!}{i!j!(k-i-j)!}\, S(i,j,k-i-j)\ .
\eeq
Gauge invariance of particle 3 is automatic. 
Note that this solution only exists for $k$ smaller (or equal) than all the spins $l_i$. Therefore,
the number of possible scattering amplitudes between 3 massless higher spin particles is $$1+{\rm min}(l_1,l_2,l_3)\ .$$
This matches the counting of conformal three-point functions of conserved tensors in $d\ge4$ (see table 1).

It is also interesting to notice the permutation symmetry properties
\beq
T_k(1,2,3)=T_k(2,3,1)=T_k(3,1,2)= (-1)^{\sum l_i} T_k(2,1,3)\,.
\eeq
In particular, this means that photons don't interact; one needs a non-abelian gauge symmetry to have a three-point function of spin 1 massless particles. 

To make further contact with the results of section \ref{sec:CE}, we can consider the case when one of the three particles is massive. In this case the analysis is simplified by going to the rest frame of the massive particle, so that we are dealing with a decay amplitude. It is also helpful to completely fix the gauge symmetry. The amplitude has to be constructed by contracting the purely spatial polarization tensors $\varepsilon_{1,2,3}$ with the spatial momentum of the decay products $\boldsymbol{p}$. We will assume that the decaying particle 3 has arbitrary spin $l$, while the massless decay products have the same spin $j$, focusing on the case $j=1,2$. Since $\varepsilon_{1,2}$ are transverse to $\boldsymbol{p}$, it's easy to construct the amplitudes:
\begin{align}
j=1:&\quad (\eps_{1}\cdot \eps_{2})(\eps_3\cdot \boldsymbol{p}^l)\,, 
\qquad
\eps_{1\mu_1}\eps_{2\mu_2} (\eps_3\cdot \boldsymbol{p}^{l-2})^{\mu_1\mu_2}\,,\\
j=2:&\quad (\eps_{1}\cdot \eps_{2})(\eps_3\cdot \boldsymbol{p}^l)\,, 
\quad
(\eps_{1}\cdot\eps_{2})_{\mu_1\mu_2}
(\eps_3\cdot \boldsymbol{p}^{l-2})^{\mu_1\mu_2}\,,
\quad
\eps_{1\mu_1\mu_2}\eps_{2\mu_3\mu_4}(\eps_3\cdot \boldsymbol{p}^{l-4})^{\mu_1\mu_2\mu_3\mu_4}\,.
\nonumber
\end{align}
This matches the number of structures found in table 1, including the symmetry/anti-symmetry of the current correlators, corresponding to parity under $\boldsymbol{p}\to-\boldsymbol{p}$. Notice that for low $l$ one runs out of indices to contract with $\boldsymbol{p}$ and the number of amplitudes is reduced, again in agreement with table 1.  

\subsection{Four Dimensions}

In four dimensions, the 5 vectors $z_1,z_2,z_3,p_1,p_2$ can not be linearly independent. 
Therefore, the determinant 
\beq
\left. \det_{1\le i,j \le 5}  (z_i\cdot z_j )\right|_{z_4=p_1 \atop z_5=p_2} 
\eeq
must vanish.
This gives the following identity:
\begin{align}
\left(\frac{1}{2} \sum_i^3 M_i^4 \right.&\left.- \sum _{i<j}^3 M^2_iM^2_j \right)
(z_1\cdot z_2 )(z_1\cdot z_3 )(z_2\cdot z_3 )\nonumber \\
&=2(z_1\cdot z_2 ) (z_1\cdot p_2 )(z_2\cdot p_3 )(z_3\cdot p_1 )^2
-M_1^2 (z_1\cdot p_2 )(z_2\cdot z_3 )
\label{detSmatrix} \nonumber
\\&\ \ \ +(M^2_1-M^2_2-M^2_3)(z_2\cdot p_3 ) (z_3\cdot p_1 )
(z_1\cdot z_2 )(z_1\cdot z_3 ) +{\rm cyclic}\,.
\end{align}
Thus, we do not need to use the structure $(z_1\cdot z_2 )(z_1\cdot z_3 )(z_2\cdot z_3 )$, and we recover precisely the counting of CFT three-point functions in three dimensions.

In the massless case, there is an even simpler relation
\begin{align}
0&=(z_2\cdot p_3)(z_3\cdot p_1)  z_1
+(z_1\cdot p_2)(z_3\cdot p_1)  z_2
+(z_2\cdot p_3)(z_1\cdot p_2)  z_3 \label{vectoridentity} \nonumber
\\&\ \ \ -(z_1\cdot z_3)(z_2\cdot p_3)  p_1
+(z_2\cdot z_3)(z_1\cdot p_2)  p_2 \,.
\end{align}
Taking the inner product with $z_1$ we re-obtain the identity (\ref{detSmatrix}) in the massless case
\begin{align}
  (z_1\cdot z_2)(z_3\cdot p_1)
+  (z_1\cdot z_3) (z_2\cdot p_3) 
+(z_2\cdot z_3)  (z_1\cdot p_2) = 0\,,
\end{align}
which relates the basic structures as
\beq
S(n_{12}+1,n_{13},n_{23})+S(n_{12},n_{13}+1,n_{23})+S(n_{12},n_{13},n_{23}+1) = 0\,,
\eeq
assuming that all $m_i=l_i-\sum_{j} n_{ij}$ are non-zero.
Therefore, we can write all structures in terms of structures with $n_{23}=0$:
\beq
S(n_{12},n_{13},n_{23})=(-1)^{n_{23}} \sum_{i=0}^{n_{23}}
\frac{n_{23}!}{i!(n_{23}-i)!}S(n_{12}+i,n_{13}+n_{23}-i,0)\,.
\label{3Dn23=0}
\eeq
This reduces the gauge invariant amplitude to
\begin{align}
T_k&=\sum_{i=0}^k \sum_{j=0}^{k-i} \frac{k!(-1)^{k-i-j} }{i!j!(k-i-j)!} 
\sum_{t=0}^{k-i-j}
\frac{(k-i-j)!}{t!(k-i-j-t)!}S(i+t,k-i-t,0)\nonumber\\
&=\sum_{i+j+t+u=k}
 \frac{k!(-1)^{t+u} }{i!j! t!u!}  S(i+t,j+u,0)\nonumber\\
&=\sum_{r+s=k}  S(r,s,0) \frac{k! }{r! s!} \sum_{i+t =r}(-1)^{t }\frac{r! }{i!  t! }  \sum_{ j+u=s} (-1)^{ u}
 \frac{s! }{ j!  u!} \nonumber\\
 &=0 
\end{align}
in general.  However, there are two special cases: $k=0$ and $k={\rm min}(l_1,l_2,l_3)$.
It is clear that $T_0$ does not vanish identically and is gauge invariant.
When $k$ takes its maximal value, equal to the smaller spin (which we choose to be $l_1$), the identity (\ref{3Dn23=0}) can not be used.
In particular, $S(i,l_1-i-1,1)$ with $i=0,1,\dots,l_1-1$ can not be written solely in terms of structures with $n_{23}=0$. The best we can do is to reduce it down to structures with $n_{23}=0$ and $n_{23}=1$. 
We conclude that in 4 dimensions there are only 2 parity even structures for the scattering amplitude of 3 massless higher spin fields.
This result agrees with the conjecture of \cite{Giombi} that there are only 2 independent structures for the three-point function of conserved tensors in CFT$_3$.

Here we only considered parity even structures. It should be possible to give an analogous discussion for parity odd structures, where we expect to find one amplitude if the spins $\{l_1,l_2,l_3\}$ satisfy the triangle inequality and zero otherwise, to match the conjecture of \cite{Giombi} in the parity odd case.

\subsection{Relation to AdS/CFT Duality}

 In the case of polynomial scattering amplitudes, we can use AdS/CFT to provide an explicit map from scattering amplitudes in $\bb{M}^{d+1}$ to CFT$_{d}$ correlators. 
 We simply construct a contact Witten diagram that connects $n$ bulk-to-boundary propagators to the local interaction vertex corresponding to the  
 $n$-particle S-matrix element. 
 This map was already explored in the case of four-point functions of scalar operators in \cite{joao,IdseJamie}.
Above, we saw that it should also extend to $n$-point functions of tensor operators.
However, when the scattering amplitude has poles describing a mediated interaction, the situation is more complicated. 
It would be very interesting to construct an explicit map from  S-matrix elements to conformal correlators that is also valid in this case.
The Mellin representation of conformal correlators \cite{Mack,joaoMellin,Fitzpatrick:2011ia,Paulos:2011ie} may be useful in this context, given its close structural analogy to scattering amplitudes.

Let us now give this map explicitly in the simplest case of  three particle scattering. 
To each S-matrix element $S(n_{12},n_{13},n_{23})$ given in (\ref{3ptSmatrices}) we can associate a cubic local interaction vertex in the Lagrangian for AdS fields given by
\beq
{\cal V}(n_{12},n_{13},n_{23})= \left(\left(\nabla_\nu\right)^{m_2} \phi_1^{\mu_1\dots \mu_{l_1}} \right)
\left(\left(\nabla_\rho\right)^{m_3}   \phi_2^{\nu_1\dots \nu_{l_2}} \right)
\left(\left(\nabla_\mu\right)^{m_1} \phi_3^{\rho_1\dots \rho_{l_3}} \right)
(g_{\mu\nu})^{n_{12}}
(g_{\mu\rho})^{n_{13}}
(g_{\nu\rho})^{n_{23}}\,,
\label{vertex}
\eeq
where Greek indices denote AdS indices.
We use a schematic notation where, for example, $(\nabla_\nu)^{m_2}$ is the covariant derivative acting $m_2$ times on the 
field $\phi_1$, with  indices contracted with the $\nu$ indices of the field $\phi_2$. The notation used in $(g_{\mu\nu})^{n_{12}}$
tells us that there are $n_{12}$ contractions of the indices of the fields $\phi_1$ and  $\phi_2$.
We recall that the integers $m_i$ are determined by the $n_{ij}$'s through the constraint $m_i + \sum_jn_{ij} = l_i$.

The AdS/CFT duality gives  an explicit rule on how to map the above interaction vertex to a correlation function of
operators dual to the fields $\phi_i$: one simply computes the Witten diagram by  replacing in 
(\ref{vertex}) the fields by their bulk-to-boundary propagators, and then integrates over the AdS interaction
point. We shall denote the bulk-to-boundary propagator from an AdS point $y$ to a boundary point $x$ by
\beq
\Pi^{\mu_1\dots \mu_l, a_1\dots a_l}(y,x)\,.
\eeq
This propagator obeys the bulk equation
\beq
\nabla^\nu \nabla_\nu \Pi^{\mu_1\dots \mu_l, a_1\dots a_l}  = \left(  \Delta (\Delta - d) - l \right) \Pi^{\mu_1\dots \mu_l, a_1\dots a_l}\,,
\label{box}
\eeq
and has vanishing divergence
\beq
\nabla_\mu\Pi^{\mu \mu_2\dots \mu_l, a_1\dots a_l} = 0 \,.
\eeq
From AdS/CFT one expects that all three-point functions can be written as a linear combination of this set of Witten diagrams.
Of course the basis of three-point functions obtained this way is not the
same basis of section \ref{sec:corr}. 
In particular, Witten diagrams give a basis of tensor structures where the constraints arising from 
operator conservation are simpler to formulate.

Let us then analyze in more detail the case of conserved spin $l$ operators. We wish to understand the constraints
imposed on the bulk interaction vertices ${\cal V}(n_{12},n_{13},n_{23})$ that arise from current conservation in the CFT side.
The boundary divergence acting on the bulk-to-boundary
propagator  of dimension $\Delta = d - l + 2$ is pure gauge, i.e.
\beq
\partial_a \, \Pi^{\mu_1\dots \mu_{l_1}, a a_2\dots a_l} =
\nabla^{( \mu_1}
\Lambda^{\mu_2\dots \mu_{l_1} ), a_2\dots a_{l}}  \,,
\label{puregauge}
\eeq
where $\Lambda$ satisfies the bulk equation (\ref{box}). 
Therefore, as expected, current conservation in the boundary becomes gauge invariance in the bulk.\footnote{In the original three graviton case of Hofman and Maldacena~\cite{Hofman} this gauge invariance was general covariance and the vertices were extracted from a generally covariant Lagrangian including the Einstein-Hilbert term and contractions of the Weyl tensor.}

Let us then look for gauge invariant  linear combinations of vertices of the type
\beq
{\cal V} = \sum_{\{n_{ij}\}} a(n_{ij}) {\cal V}(n_{ij})\,.
\label{VertexExpansion}
\eeq
Suppose that we consider the field
$\phi^{\mu_1\dots \mu_l}_1= \nabla^{(\mu_1} \Lambda^{\mu_2\dots \mu_l)}$ to be pure gauge.
After some integrations by parts, and using the equations of motion,
the vertex ${\cal V} (n_{12},n_{13},n_{23})$
transforms to
\beq
\delta_1{\cal V} (n_{ij}) =    \gamma \left( l_1-n_{12}-n_{13}  \right) \tilde{\cal V} (n_{ij})  + n_{12} \tilde{\cal V} (n_{12}-1,n_{13},n_{23}) 
-n_{13}  \tilde{\cal V} (n_{12},n_{13}-1,n_{23}) \,,
\eeq
where
\beq
  \gamma  = \frac{1}{2}\left(  \mu_2^2 - \mu_1^2 - \mu_3^2\right)\,,\ \ \ \ \ \ \ \ \ \ \ \mu^2_i=  \Delta_i (\Delta_i - d) - l_i\,.
\eeq
Note that here $ \tilde{\cal V}$ denotes the vertex introduced in (\ref{vertex}) with  $\phi_1$ replaced by 
the gauge tensor $\Lambda$ of spin $l_1-1$. 
This equation is the direct analogue of (\ref{gaugecondition1}) in flat space. The only difference is the appearance of an extra term proportional to the mass squared of the higher spin gauge fields in AdS.
We conclude that gauge invariance imposes the constraint 
\beq
\gamma \left( l_1-n_{12}-n_{13}  \right) a (n_{ij})  + \left( n_{12}+ 1\right) a (n_{12}+1,n_{13},n_{23})  - \left( n_{13}+ 1\right)  a(n_{12},n_{13}+1,n_{23})=0 \,,
\eeq
on the coefficients of the expansion (\ref{VertexExpansion}).
Imposing gauge invariance on $\phi_2$ and $\phi_3$ produces similar equations.  

\section{Summary and Conclusions}
\label{sec:concl} 

With the formalism developed in this paper, the kinematical constraints arising from conformal invariance 
can be implemented for symmetric traceless operators of arbitrary spin almost as easily as for  scalar operators. 
On the next page, we  briefly summarize the basic rules for the more pragmatic reader.

\begin{figure}
\fbox {
    \parbox{\linewidth}{
    
    \begin{center}
    \Large{\bf Summary}
\end{center}

\begin{itemize}
\item{\bf Embedding space}

The natural habitat for conformal field theories is the light cone of the origin of $\bM^{d+2}$.
\CG~Lorentz transformations of the light rays generate conformal transformations. 
The usual flat physical space  $\bR^d$ can be obtained by projecting into the Poincar\'e section 
of the light cone
\beq
P_x= (P^+,P^-,P^a) =( 1, x^2, x^a)\,.
\eeq

\item{\bf Primary fields}

Primary fields of dimension $\Delta$ and spin $l$ are encoded into a field $F(P,Z)$, polynomial in the polarization vector $Z$, such that
\beq
F(\lambda P; \alpha Z + \beta P ) = \lambda^{-\Delta} \alpha^l F(P;Z)\,.
\label{EmbField}
\eeq
The usual tensor form of the operator on $\bR^d$ is obtained from
\beq
f_{a_1\dots a_l}(x) = \frac{1}{l! (h-1)_l}  D_{a_1} \cdots D_{a_l} F(P_x;Z_{z,x})\,,
\eeq
where $D_a$ is the differential operator defined in (\ref{Tod}) and $Z_{z,x}=(0,2x\cdot z, z^a)$.

\item{\bf Correlators}

The most general form of the correlator 
\beq
\left\langle F_1(P_1;Z_1) \cdots  F_n(P_n;Z_n) \right\rangle
\eeq
compatible with conformal invariance is a linear combination of homogeneous polynomials of degree $l_i$ in each $Z_i$, each
constructed by multiplying the basic building blocks $V_{i,jk}$ and $H_{ij}$ given in (\ref{V_i,jk}) and (\ref{H_ij}).
The $P_i$ dependence is then constrained by the scaling in (\ref{EmbField}).

\item{\bf Conserved fields}

A spin $l$ primary field of dimension $\Delta= d- 2 +l$ obeys the conservation equation
\beq
\left(\partial_{P} \cdot D \right) F(P;Z) = 0\,,
\eeq
where $D$ is the differential operator defined in (\ref{EmbTod}).  This condition
generates additional constraints on the correlators of conserved fields that can be easily implemented.
\end{itemize}

    }
}
\end{figure}

The focus of this paper was to develop the formalism, postponing applications to the near future~\cite{Costa:2011dw}. We were careful to establish connections with previous work and exemplify the strength of the method by rederiving many known results. For example, we have established a one-to-one correspondence with the general three-point function analysis of Mack \cite{Mack} and of Osborn and Petkou \cite{OP}, as well as with recent work on the three dimensional case in \cite{Giombi}.

We have also presented several new results, interesting in their own right. For example, we reduced the problem of counting conformal three-point functions of operators with spin to the simple combinatorial problem depicted in Figure \ref{Diagram3pt}, which we solved in closed form in Eq.~(\ref{N3pt}).
For spin 1 currents and the stress tensor, we studied how conservation leads to a reduction in the number of three-point functions with an arbitrary spin $l$ primary. We have also discussed a general rule for counting the number of three-point functions in terms of flat space S-matrices in $d+1$ dimensions.
Using this rule, we conjecture that the number of independent tensor structures 
for three-point functions of conserved tensors in $d\ge4$ is given by $1+{\rm min}(l_1,l_2,l_3)$.
In three dimensions, the number of structures is reduced to 2 as claimed in \cite{Giombi}.

In this paper we have been dealing with correlators of bosonic fields, but it should be pointed out that the embedding formalism can be also developed for fermion correlators \cite{Weinberg}. Finally, although we have limited the discussion to the symmetric traceless primaries, it should not be too difficult to extend the formalism to anti-symmetric fields or fields of mixed symmetry, using polynomials in Grassmann variables.


\bigskip

\bigskip


\begin{center} 
{\bf Acknowledgements} 
\end{center}
M.C. and J.P. are grateful to Lorenzo Cornalba for many discussions in the early stages of this project.  We thank Diego Hofman for discussions related to Section~\ref{sec:S}.  D.P. also thanks David Simmons-Duffin for many related discussions. This work was partially funded by the research 
grants PTDC/FIS/099293/2008, CERN/FP/116358/2010 and by the UNIFY IRSES Marie Curie network.
\emph{Centro de F\'{i}sica do Porto} is partially funded by FCT under grant PEst-OE/FIS/UI0044/2011.
The work of S.R. was supported in part by
the European Program ``Unification in the LHC Era",
contract PITN-GA-2009-237920 (UNILHC) and by the National Science Foundation under Grant No. NSF PHY05-51164. S.R. is grateful to the Perimeter Institute and to KITP, Santa Barbara, for hospitality.  D.P. is supported in part by the Harvard Center for the Fundamental Laws of Nature and by NSF grant PHY-0556111.
Research at the Perimeter Institute is supported in part by the Government of Canada through NSERC and by the Province of Ontario through the Ministry of Research \& Innovation. We thank Fernando Rejon-Barrera for pointing out a misprint in Eq.~(\ref{A.9}).


\appendix

\section{Three-Point Function for (Spin 2)-(Spin 2)-(Spin $l$)}

\label{app:a}
In this appendix we will apply the formalism developed in section~\ref{sec:cons} to the case of a three-point function between the spin 2 stress tensor $T_{ab}$ at $x_1$ and $x_2$ and a  dimension $\Delta$ operator of spin $l$ at $x_3$,
\beq
\langle  T_{ab}(x_1) T_{cd}(x_2)  {\cal O}_{e_1\cdots e_l}(x_3) \rangle .
\label{ttl}
\eeq
When $l$ is even, the embedding function (prior to imposing the conservation constraints) has 10 possible structures with coefficients $\alpha_a$:
\beq
\tilde G(\{P_i; Z_i\}) = 
\frac{ \sum\limits_{a=1}^{10} \alpha_a A_a(V_i,H_{ij}) }{\left(  P_{12} \right)^{d+2-\frac{\Delta+l}{2} }\left(  P_{13}\right)^{\frac{\Delta+l}{2} }\left( P_{23} \right)^{\frac{\Delta+l}{2} }} ,
\label{TTL}
\eeq
where the structures symmetric under exchanging $\{P_1,Z_1\}$ with $\{P_2,Z_2\}$ are given by
\beq
A_a(V_i,H_{ij}) = \left( 
\begin{array}{c}
V_1^2 V_2^2 V_3^l \\
(H_{13} V_2^2 V_1 + H_{23} V_1^2 V_2) V_3^{l-1} \\
H_{12} V_1 V_2 V_3^l \\
(H_{13} V_2 + H_{23} V_1) H_{12} V_3^{l-1} \\
H_{13} H_{23} V_1 V_2 V_3^{l-2} \\
H_{12}^2 V_3^l \\
(H_{13}^2 V_2^2 + H_{23}^2 V_1^2) V_3^{l-2} \\
H_{12} H_{23} H_{13} V_3^{l-2} \\
(H_{13} H_{23}^2 V_1 + H_{23} H_{13}^2 V_2) V_3^{l-3} \\
H_{13}^2 H_{23}^2 V_3^{l-4} \\
\end{array}
\right) .
\eeq
We can then compute the divergence at $P_1$ and drop terms of $O(Z_1^2, Z_1\cdot P_1)$ to obtain
\beq
(\partial_{P_1}\cdot D_{Z_1})\,\tilde G \to \frac{ \sum\limits_{a=1}^{8} \beta_a B_a(V_i,H_{ij}) }{\left(  P_{12} \right)^{d+2-\frac{\Delta+l}{2} }\left(  P_{13}\right)^{\frac{\Delta+l}{2} }\left( P_{23} \right)^{\frac{\Delta+l}{2} }} ,
\eeq
where we have chosen the basis of structures
\beq
B_a(V_i,H_{ij}) = \left( 
\begin{array}{c}
V_1 V_2^2 V_3^l \\
H_{13} V_2^2 V_3^{l-1} \\
H_{23} V_1 V_2 V_3^{l-1} \\
H_{12} V_2 V_3^{l} \\
H_{13} H_{23} V_2 V_3^{l-2} \\
H_{12} H_{23} V_3^{l-1} \\
H_{23}^2 V_1 V_3^{l-2} \\
H_{13} H_{23}^2 V_3^{l-3} \\
\end{array}
\right) ,
\eeq
and the coefficients $\beta_a$ are given by
\begin{align}
\beta_1 &= \alpha_1 \Bigl( 2-l+\Delta-d(1-d+\Delta) \Bigr) + \alpha_2 \Bigl(-2+l-\Delta+\frac12 d (2-2d-l+\Delta) \Bigr)  \nonumber\\
& \qquad + \alpha_3 \Bigl( -2+l-\Delta+\frac12 d (2+l+\Delta) \Bigr) + 2 \alpha_4 \Bigl( 2-d-l+\Delta\Bigr) , \\
\beta_2 &= - \alpha_1 l + \frac12 \alpha_2 \Bigl( d^2+2 l -d \Delta  \Bigr) +\alpha_3 l + \frac12 \alpha_4 \Bigl((d-4)l +d \Delta \Bigr) + \alpha_7 d (-2d-l+\Delta) , \\
\beta_3 &= \alpha_2 \Bigl( 2+d^2-l+\Delta-d(2+\Delta) \Bigr) + \frac12 \alpha_3 d l + \frac12 \alpha_4 d (-2+l+\Delta) \nonumber \\
& \qquad + \alpha_5 \Bigl( -2+l-\Delta+\frac12 d (4-2d-l+\Delta) \Bigr) - 2 \alpha_8 (-2+d+l-\Delta) , \\
\beta_4 &= 2 \alpha_1 -2 \alpha_2 +\frac12 \alpha_3 \Bigl(-4+d^2-d \Delta \Bigr) + \alpha_4 \Bigl(4-\frac12 d (2d+l-\Delta) \Bigr) + \alpha_6 d (l+\Delta) \label{A.9},\\
\beta_5 &= -\alpha_2 (l-1) +\frac12 \alpha_4 d (l-1)+\frac12 \alpha_5 \Bigl(-2+d^2+2l-d\Delta \Bigr) - 2 \alpha_7 d \nonumber \\
& \qquad +\frac12 \alpha_8 \Bigl( 4-4l+d(-2+l+\Delta) \Bigr) + \alpha_9 d (2-2d-l+\Delta) , \\
\beta_6 &= \alpha_2 + \frac12 \alpha_4 d (-1+d-\Delta) -\alpha_5 + \alpha_6 d l +\alpha_8 \Bigl( 2+\frac12 d(2-2d-l+\Delta) \Bigr) , \\
\beta_7 &= \frac12 \alpha_4 d (l-1) - \frac12 \alpha_5 d + \alpha_7 \Bigl( 2+d^2-l+\Delta-d(1+\Delta) \Bigr) +\alpha_8 \Bigl(2-d-l+\Delta \Bigr) \nonumber\\
& \qquad -\frac12 \alpha_9 (d-2) \Bigl(-2+2d+l-\Delta \Bigr) , \\
\beta_8 &= -\alpha_7 (l-2) +\frac12 \alpha_8 (d-2)(l-2) +\frac12 \alpha_9 \Bigl(d^2+2(l-2)-d(2+\Delta) \Bigr) \nonumber \\
& \qquad + \alpha_{10} d \Bigl(4-2d-l+\Delta \Bigr) .
\end{align}
Setting each of these coefficients to zero would na\"ively reduce the number of structures from 10 down to 2.  However, one of the equations is linearly dependent due to the relation
\begin{align}
0 &= 2 \beta_1 l  + 2 \beta_2 (d^2-l+\Delta-d \Delta)  - \beta_3 (2l+d(2+d-\Delta))  + \beta_4 (d-2) l  \nonumber \\
& \qquad -\beta_5(d-2)(2d+l-\Delta)   - \beta_6 (d-2) (l+\Delta)  + 2\beta_7 d(2d+l-\Delta),
\end{align}
so the number of independent structures is actually reduced from 10 down to 3.

Next let us consider the case that $l$ is odd. In this case there are initially 4 possible structures invariant under exchanging $\{P_1,Z_1\}$ with $\{P_2,Z_2\}$:
\beq
\tilde G(\{P_i; Z_i\}) = 
\frac{ \sum\limits_{a=1}^{4} \gamma_i G_a(V_i,H_{ij}) }{\left(  P_{12} \right)^{d+2-\frac{\Delta+l}{2} }\left(  P_{13}\right)^{\frac{\Delta+l}{2} }\left( P_{23} \right)^{\frac{\Delta+l}{2} }} ,
\label{TTL}
\eeq
with
\beq
G_a(V_i,H_{ij}) = \left( 
\begin{array}{c}
(H_{13} V_2^2 V_1  - H_{23} V_1^2 V_2) V_3^{l-1} \\
(H_{13} V_2 - H_{23} V_1) H_{12} V_3^{l-1} \\
(H_{13}^2 V_2^2 - H_{23}^2 V_1^2 ) V_3^{l-2} \\
(H_{23}^2 H_{13} V_1 -  H_{13}^2 H_{23} V_2) V_3^{l-3}
\end{array}
\right) .
\eeq
Then computing the divergence at $P_1$ and dropping the usual terms gives 
\beq
(\partial_{P_1}\cdot D_{Z_1})\,\tilde G \to \frac{ \sum\limits_{a=1}^{8} \delta_a B_a(V_i,H_{ij}) }{\left(  P_{12} \right)^{d+2-\frac{\Delta+l}{2} }\left(  P_{13}\right)^{\frac{\Delta+l}{2} }\left( P_{23} \right)^{\frac{\Delta+l}{2} }} ,
\eeq
where the coefficients $\delta_i$ are given by
\begin{align}
\delta_1 &= \frac12 \gamma_1 \Bigl(  (d-2) (2-l+\Delta)-2 d^2 \Bigr) + 2 \gamma_2 \Bigl(2-d-l+\Delta \Bigr), \\
\delta_2 &= \frac12 \gamma_1 \Bigl(d^2+2l-d \Delta \Bigr) + \frac12 \gamma_2 \Bigl((d-4)l + d \Delta \Bigr) +\gamma_3 d \Bigl(-2d-l+\Delta \Bigr), \\
\delta_3 &= \gamma_1 \Bigl(-2+l-\Delta-d(d-\Delta)\Bigr) +\gamma_2 \Bigl(2(2-l+\Delta)-\frac12 d(2+l+\Delta) \Bigr), \\
\delta_4 &= -2\gamma_1 +\gamma_2 \Bigl( 4-\frac12 d(2d+l-\Delta)\Bigr), \\
\delta_5 &= \gamma_1 (l-1) + \frac12 \gamma_2 (d-4)(l-1) -2 \gamma_3 d + \gamma_4 d\Bigl(-2+2d+l-\Delta \Bigr), \\
\delta_6 &= -\gamma_1 +\gamma_2 \Bigl( 2-\frac12 d(1+d-\Delta) \Bigr), \\
\delta_7 &= -\frac12 \gamma_2 d (l-1) +\gamma_3 \Bigl(-2+l-\Delta+d(1-d+\Delta)\Bigr) \nonumber \\
& \qquad +\frac12 \gamma_4 (d-2) (2-2d-l+\Delta), \\
\delta_8 &=\gamma_3 (l-2) +\gamma_4 \Bigl( -2+l +\frac12 d(2+d-\Delta) \Bigr) .
\end{align}
Setting each of these coefficients to zero, it is straightforward to verify that there are precisely 4 linearly independent constraints, forcing $\gamma_1 = \gamma_2 = \gamma_3 = \gamma_4 = 0$.  Thus, an odd $l$ operator cannot appear in the OPE of the stress tensor with itself.
 
  

\begin{thebibliography}{99} 


\bibitem{FGG-paper}
S.~Ferrara, A.~F.~Grillo and R.~Gatto,
``Tensor Representations of Conformal Algebra and Conformally Covariant Operator Product Expansion,''
Annals Phys.\ {\bf 76} (1973) 161.

\bibitem {pol}A.~M.~Polyakov, \textquotedblleft Nonhamiltonian approach to
conformal quantum field theory,\textquotedblright%
\ Zh.\ Eksp.\ Teor.\ Fiz.\ \textbf{66} (1974) 23 [JETP \textbf{39} (1974) 10�18].

\bibitem {us}{ R.~Rattazzi, V.~S.~Rychkov, E.~Tonni and A.~Vichi,
\textquotedblleft Bounding scalar operator dimensions in 4D
CFT,\textquotedblright\ JHEP \textbf{0812}, 031 (2008),
\href{http://arxiv.org/abs/0807.0004}{arXiv:0807.0004} [hep-th];\\
}
V.~S.~Rychkov and A.~Vichi, \textquotedblleft Universal
Constraints on Conformal Operator Dimensions,\textquotedblright%
\ Phys.\ Rev.\ D \textbf{80}, 045006 (2009),
\href{http://arxiv.org/abs/0905.2211}{arXiv:0905.2211} [hep-th];\\
F.~Caracciolo and V.~S.~Rychkov, \textquotedblleft Rigorous
Limits on the Interaction Strength in Quantum Field Theory,\textquotedblright%
\ Phys.\ Rev.\ D \textbf{81}, 085037 (2010),
\href{http://arxiv.org/abs/0912.2726}{arXiv:0912.2726} [hep-th];\\
R.~Rattazzi, S.~Rychkov and A.~Vichi,
  ``Central Charge Bounds in 4D Conformal Field Theory,''
  Phys.\ Rev.\  D {\bf 83}, 046011 (2011),
  \hhref{1009.2725} [hep-th];\\
  ``Bounds in 4D Conformal Field Theories with Global Symmetry,''
  J.\ Phys.\ A  {\bf 44}, 035402 (2011),
  \hhref{1009.5985} [hep-th];\\
 A.~Vichi,
``Improved Bounds for CFT's with Global Symmetries,''
\hhref{1106.4037} [hep-th].

\bibitem{joao}
  I.~Heemskerk, J.~Penedones, J.~Polchinski and J.~Sully,
  ``Holography from Conformal Field Theory,''
  JHEP {\bf 0910}, 079 (2009),
  \hhref{0907.0151} [hep-th].
  
 \bibitem{david} 
D.~Poland and D.~Simmons-Duffin,
  ``Bounds on 4D Conformal and Superconformal Field Theories,''
  JHEP {\bf 1105}, 017 (2011),
 \hhref{1009.2087} [hep-th]; \\
  D.~Poland, D.~Simmons-Duffin, A.~Vichi,
  ``Carving Out the Space of 4D CFTs,''
   \hhref{1109.5176} [hep-th].
 
\bibitem{Costa:2011dw}
  M.~S.~Costa, J.~Penedones, D.~Poland and S.~Rychkov,
  ``Spinning Conformal Blocks,''
  \hhref{1109.6321} [hep-th].
 
\bibitem{MackMellin}
  G.~Mack,
  ``D-independent representation of Conformal Field Theories in D dimensions via transformation to auxiliary Dual Resonance Models. Scalar amplitudes,''
    \hhref{0907.2407} [hep-th].

\bibitem{joaoMellin}
  J.~Penedones,
  ``Writing CFT correlation functions as AdS scattering amplitudes,''
  JHEP {\bf 1103 } (2011)  025,
  \hhref{1011.1485} [hep-th].  

\bibitem{Fitzpatrick:2010zm}
  A.~L.~Fitzpatrick, E.~Katz, D.~Poland, D.~Simmons-Duffin,
  ``Effective Conformal Theory and the Flat-Space Limit of AdS,''
  JHEP {\bf 1107}, 023 (2011).
  \hhref{1007.2412} [hep-th].

\bibitem{Fitzpatrick:2011ia}
  A.~L.~Fitzpatrick, J.~Kaplan, J.~Penedones, S.~Raju and B.~C.~van Rees,
  ``A Natural Language for AdS/CFT Correlators,''
  \hhref{1107.1499} [hep-th].

\bibitem{Paulos:2011ie}
  M.~F.~Paulos,
  ``Towards Feynman rules for Mellin amplitudes in AdS/CFT,''
  \hhref{1107.1504} [hep-th].

\bibitem{Suvrat}
    S.~Raju,
  ``BCFW for Witten Diagrams,''
  Phys.\ Rev.\ Lett.\ {\bf 106} (2011) 091601,
    \hhref{1011.0780} [hep-th];\\
    ``Recursion Relations for AdS/CFT Correlators,''
  Phys.\ Rev.\  {\bf D83 } (2011)  126002,
  \hhref{1102.4724} [hep-th].

\bibitem{BCFW}
  R.~Britto, F.~Cachazo, B.~Feng, E.~Witten,
  ``Direct proof of tree-level recursion relation in Yang-Mills theory,''
  Phys.\ Rev.\ Lett.\  {\bf 94 } (2005)  181602,
  \hhref{hep-th/0501052}.
 
  \bibitem{Dirac}
  P.~A.~M.~Dirac,
  ``Wave equations in conformal space,''
  Annals Math.\  {\bf 37}, 429 (1936).
  
\bibitem{Mack:1969rr}
  G.~Mack, A.~Salam,
  ``Finite component field representations of the conformal group,''
  Annals Phys.\  {\bf 53}, 174-202 (1969).

\bibitem{Boulware}
D.~G.~Boulware, L.~S.~Brown and R.~D.~Peccei,
``Deep-Inelastic Electroproduction and Conformal Symmetry,''
Phys.\ Rev.\ D {\bf 2} (1970) 293.

 \bibitem{FGG-book} 
 S.~Ferrara, A.~F.~Grillo and R.~Gatto,
``Conformal algebra in space-time and operator product expansion,''
Springer Tracts Mod.\ Phys.\ \textbf{67}, 1-64 (1973).   

\bibitem{CCP}
  L.~Cornalba, M.~S.~Costa and J.~Penedones,
  ``Deep Inelastic Scattering in Conformal QCD,''
  JHEP {\bf 1003}, 133 (2010),
 \hhref{0911.0043} [hep-th].

\bibitem{Weinberg}
  S.~Weinberg,
  ``Six-dimensional Methods for Four-dimensional Conformal Field Theories,''
  Phys.\ Rev.\  D {\bf 82}, 045031 (2010),
 \hhref{1006.3480} [hep-th].
  
\bibitem {DF}P.~Di Francesco, P.~Mathieu and D.~Senechal, \textquotedblleft
Conformal Field Theory,\textquotedblright\
New York, USA: Springer (1997) 890pp.

\bibitem{Bars} 
  I.~Bars,
  ``Two time physics in field theory,''
  Phys.\ Rev.\  {\bf D62}, 046007 (2000),
 \hhref{hep-th/0003100}. 
 
  \bibitem{3-pt}
  A.~M.~Polyakov,
  ``Conformal symmetry of critical fluctuations,''
  JETP Lett.\  {\bf 12}, 381-383 (1970).
  
 \bibitem{SW} E.\ M.\ Stein and G.\ Weiss,
``Introduction to Fourier Analysis on Euclidean Spaces", Princeton Univ. Press, 1971, 312pp
  
  \bibitem{Guth}
  A.~H.~Guth, D.~E.~Soper,
  ``Short Distance Behavior of the Bethe-Salpeter Wave Function,''
  Phys.\ Rev.\  {\bf D12}, 1143 (1975).
  
  \bibitem{Todorov}
  V.~K.~Dobrev, V.~B.~Petkova, S.~G.~Petrova and I.~T.~Todorov,
  ``Dynamical Derivation Of Vacuum Operator Product Expansion In Euclidean
  Conformal Quantum Field Theory,''
  Phys.\ Rev.\  D {\bf 13}, 887 (1976).


\bibitem{Belitsky:2007jp}
A.~V.~Belitsky, J.~Henn, C.~Jarczak, D.~Mueller and E.~Sokatchev,
``Anomalous Dimensions of Leading Twist Conformal Operators,''
Phys.\ Rev.\ D {\bf 77} (2008) 045029,
\hhref{0707.2936} [hep-th].


\bibitem{Waldron}
M.~Grigoriev and A.~Waldron,
``Massive Higher Spins from BRST and Tractors,''
Nucl.\ Phys.\ B {\bf 853} (2011) 291,
\hhref{1104.4994} [hep-th].


\bibitem{Todorov-book}
  V.~K.~Dobrev, G.~Mack, V.~B.~Petkova, S.~G.~Petrova, I.~T.~Todorov,
  ``Harmonic Analysis on the n-Dimensional Lorentz Group and Its Application to Conformal Quantum Field Theory,''
  Berlin 1977, 280p.
  
\bibitem{Mack}{ G.~Mack, \textquotedblleft Convergence Of Operator Product
Expansions On The Vacuum In Conformal Invariant Quantum Field
Theory,\textquotedblright%
\ \href{http://projecteuclid.org/DPubS?service=UI&version=1.0&verb=Display&handle=euclid.cmp/1103900641}{Commun.\ Math.\ Phys.\ \textbf{53}%
, 155 (1977)}.
}  
  
\bibitem{Zaikov}
G.~M.~Sotkov and R.~P.~Zaikov,
``Conformal Invariant Two Point and Three Point Functions for Fields with Arbitrary Spin,''
Rept.\ Math.\ Phys.\ {\bf 12} (1977) 375.
  
\bibitem{OP}
  H.~Osborn, A.~C.~Petkou,
  ``Implications of conformal invariance in field theories for general dimensions,''
  Annals Phys.\  {\bf 231 } (1994)  311-362,
\hhref{hep-th/9307010}.
    
\bibitem{MP}
J.~M.~Maldacena and G.~L.~Pimentel,
``On Graviton Non-Gaussianities during Inflation,''
\hhref{1104.2846} [hep-th].

\bibitem{Giombi}
  S.~Giombi, S.~Prakash, X.~Yin,
  ``A Note on CFT Correlators in Three Dimensions,''
  \hhref{1104.4317} [hep-th].
  
\bibitem{unitarity4D}S.~Ferrara, R.~Gatto and A.~F.~Grillo, {}``Positivity
Restrictions On Anomalous Dimensions," Phys.\ Rev.\ D
\textbf{9}, 3564 (1974);\\ 
G.~Mack, {}``All Unitary Ray Representations Of The Conformal Group
SU(2,2) With Positive Energy," Commun.\ Math.\ Phys.\ \textbf{55},
1 (1977). 

\bibitem{unitarity} R.R. Metsaev, ``Massless mixed symmetry bosonic free fields in 
d-dimensional anti-de Sitter space-time,'' Phys. Lett  B354, 78-84 (1995);\\
S.~Minwalla, {}``Restrictions imposed
by superconformal invariance on quantum field theories,'' Adv.\ Theor.\ Math.\ Phys.\ \textbf{2},
781-846 (1998), \hhref{hep-th/9712074}.

\bibitem{Schreier}
E.~J.~Schreier,
``Conformal Symmetry and Three-Point Functions,''
Phys.\ Rev.\ D {\bf 3} (1971) 980.

\bibitem{Hofman}
D.~M.~Hofman and J.~Maldacena,
``Conformal Collider Physics: Energy and Charge Correlations,''
JHEP {\bf 0805} (2008) 012,
\hhref{0803.1467} [hep-th].

\bibitem{Metsaev}
R.~R.~Metsaev and A.~A.~Tseytlin,
``Curvature Cubed Terms in String Theory Effective Actions,''
Phys.\ Lett.\ B {\bf 185} (1987) 52.

\bibitem{IdseJamie}
  I.~Heemskerk, J.~Sully,
  ``More Holography from Conformal Field Theory,''
  JHEP {\bf 1009 } (2010)  099,
 \hhref{1006.0976} [hep-th].
  
\end{thebibliography}
\end{document}